\documentclass[nofootinbib,prd,11pt]{revtex4}
\usepackage{graphicx}

\begin{document}


\title{ Neutrino Masses and Mixings 
  in a Minimal SO(10) Model}
\author{ K.S. Babu}
\email{babu@okstate.edu}
\affiliation{ 
Department of Physics, Oklahoma Center for High Energy Physics,
Oklahoma State University,  Stillwater, OK
74078, USA}
\author{ C. Macesanu}
\email{cmacesan@physics.syr.edu}
\affiliation{
Department of Physics, Oklahoma Center for High Energy Physics,
Oklahoma State University,  Stillwater, OK
74078, USA}
\affiliation{ 
Department of Physics,
Syracuse University, Syracuse, NY 13244-1130, USA }

\begin{abstract}

We consider a minimal formulation of $SO(10)$ Grand Unified Theory
wherein all the fermion masses arise from Yukawa couplings involving one $\overline{\bf 126}$
and one {\bf 10} of Higgs multiplets. It has recently been recognized that such theories can
explain, via the type--II seesaw mechanism, the large $\nu_\mu - \nu_\tau$ mixing  as a consequence of
$b-\tau$ unification at the GUT scale. In this picture, however, the CKM phase $\delta$ lies
preferentially in the second quadrant, in contradiction with
experimental measurements. We revisit this minimal model and show that the conventional
type--I seesaw mechanism generates phenomenologically viable neutrino masses and mixings,
while being consistent with CKM CP violation.  We also present improved fits in the
type--II seesaw scenario and suggest fully consistent fits in a mixed scenario.
\end{abstract}

\begin{flushright}
OSU-HEP-05-7\\
SU-4252-806
\end{flushright}

\maketitle

\newcommand{\beq}[1] {\begin{equation}\label{#1} }
\newcommand{\eeq} {\end{equation} }

\newcommand{\bea}[1]{\begin{eqnarray}\label{#1} }
\newcommand{\eea}{\end{eqnarray}}

\newcommand{\al}{\alpha}
\newcommand{\eps}{\epsilon}
\newcommand{\om}{\omega}
\newcommand{\del}{\delta}
\newcommand{\Del}{\Delta}
\newcommand{\lam}{\lambda}
\newcommand{\sig}{\sigma}

\newcommand{\mm}{m_{\mu}}
\newcommand{\mtau}{m_{\tau}}

\newcommand{\tme}{\tilde{m_e}}
\newcommand{\tmm}{\tilde{m_{\mu}}}
\newcommand{\tmt}{\tilde{m_t}}
\newcommand{\tmb}{\tilde{m_b}}
\newcommand{\ta}{\tilde{a}}
\newcommand{\tb}{\tilde{b}}
\section{Introduction}

Grand Unified Theories (GUT) provide a natural framework to
understand the properties of fundamental particles such as their
charges and masses. GUT models based on $SO(10)$ gauge symmetry
have a number of particularly appealing features. All the fermions
in a family fit in a single 16--dimensional spinor multiplet of
$SO(10)$. In order to complete this multiplet, a right--handed
neutrino field is required, which would pave the way for the
seesaw mechanism which explains the smallness of left--handed
neutrino masses. $SO(10)$ contains $SU(5)$ and the left--right
symmetric Pati--Salam symmetry group $SU(4)_C \times SU(2)_L
\times SU(2)_R$ as subgroups, both with very interesting
properties from a phenomenological perspective. With low energy
supersymmetry, $SO(10)$  and $SU(5)$ models also lead remarkably
to the unification of the three Standard Model gauge couplings at
a scale $M_{\rm GUT} \sim 2 \times 10^{16}$ GeV.

In grand unified theories, the gauge sector and the fermionic matter sector are generally
quite simple. However, the same is not true of the Higgs sector.  Since the larger symmetry
needs to be broken down to the Standard Model, generally
one needs to introduce a large number of Higgs multiplets, with different
symmetry properties under gauge transformations. If all of these Higgs fields
couple to the fermion sector, one would lose much of the predictive
power of the theory in the masses and mixings of quarks and leptons,
and so also one of the attractive aspects of GUTs.

Of interest then are the so--called minimal unification theories, in
which only a small number of Higgs multiplets couple to the fermionic
sector. One such realization is the minimal $SO(10)$ GUT \cite{babu} in which
only one {\bf 10} and one $\overline{\bf 126}$ of Higgs fields couple to the fermions.
These two Higgs fields are responsible for giving masses
to all the fermions of the theory, including large Majorana masses to the right--handed
neutrinos. This model is minimal in the following sense.  The fermions belong to
the ${\bf 16}$ of $SO(10)$, and the fermion bilinears are given by
 ${\bf 16} \times {\bf 16} = {\bf 10}_s +{\bf 120}_a +
{\bf 126}_s$.  Thus  {\bf 10}, {\bf 120} and $\overline{\bf 126}$
Higgs fields can have renormalizable Yukawa couplings.  If only
one of these Higgs fields is employed, there would be no family
mixings, so two is the minimal set.  The $\overline{\bf 126}$ has
certain advantages. It contains a Standard Model singlet field and
so can break $SO(10)$ down to $SU(5)$, changing the rank of the
group.  
 Its Yukawa couplings to the fermions also provide large
Majorana masses to the right--handed neutrinos leading to the
seesaw mechanism.  It was noted in Ref. \cite{babu} that due to
the cross couplings between the $\overline{\bf 126}$ and the {\bf
10} Higgs fields, the Standard Model doublet fields contained in
the $\overline{\bf 126}$ will acquire vacuum expectation values
(VEVs) along with the VEVs of the Higgs doublets from the ${\bf 10}$.
The $\overline{\bf 126}$ Yukawa coupling matrix will then
contribute both to the Dirac masses of quarks and leptons, as well
as to the Majorana masses of the right--handed neutrinos.

It is not difficult to realize that this minimal model is highly constrained in explaining
the fermion masses and mixings.  There are two complex symmetric Yukawa coupling matrices, one of
which can be taken to be real and diagonal without loss of generality.  These matrices have 9 real
parameters and six phases.  The mass matrices also depend on two ratios of VEVs, leading to 11 magnitudes
and six phases in total in the quark and lepton sector, to be compared with the 13 observables (9 masses,
3 CKM mixings and one CP phase).  Since the phases are constrained to be between $-1$ and $+1$, this
system does provide restrictions.  More importantly, once a fit is found for the charged fermions,
the neutrino sector is fixed in this model.  It is not obvious at all that including the neutrino
sector the model can be phenomenologically viable.

Early  analyses \cite{babu,lavoura} found that just fitting the
lepton-quark sector is highly constraining. Also, this fitting has
been found to be highly nontrivial (in terms of complexity);
therefore these analyses were done in the limit when the phases
involved are either zero or $\pi$. In such a framework, one finds
that the parameters of the models are more or less determined by
the  fit to  the lepton-quark sector (the quark masses themselves
are not known with great precision, so there is still some room
for small variations of the parameters). As a consequence, one
could more or less predict the neutrino masses and mixings;
however, since neutrino data was rather scarce at the time, one
could not impose meaningful constraints on the minimal $SO(10)$
model from these predictions.

 In view of the new information on the neutrino sector gathered in the
past few years \cite{solar,atm,chooz},
one should ask if this model  is still
consistent with experimental data.
Interest in the study of this model has also been reawakened
by the observation that $b-\tau$ unification at GUT scale implies
large (even close to maximal) mixing in the 2-3 sector of
the neutrino mass matrix \cite{bajc}, provided that the dominant
contribution to the neutrino mass is from  type--II seesaw.
 There has been a number
of recent papers studying the minimal $SO(10)$ using varying approaches:
some analytical, concentrating on the 23 neutrino sector \cite{bajc,Bajc2},
some numerical, either in the approximation that the phases involved
in reconstructing the lepton sector are zero 
\cite{fukuyama01,fukuyama02,fukuyama,moha1}, or
taking these phases into account \cite{moha2,Dutta1}. The conclusions
of these analyses seem to be that the minimal $SO(10)$  cannot
account by itself for the observed neutrino sector (although it comes
pretty close). However, one might restore agreement with the neutrino data if
one slightly modifies the minimal $SO(10)$; for example,
one can set the quark sector CKM phase to lie in the second quadrant, and
rely on new contributions  from the SUSY breaking
sector in order to explain data on quark CP violation \cite{moha2};
or one might add higher dimensional operators to the theory \cite{moha2,Dutta1},
or even another Higgs multiplet (a {\bf 120}) which will serve as small
perturbation to fermion masses \cite{Dutta2,Bertolini,Bertolini:2005}.

In this paper we propose to revisit the analysis for the {\it minimal} $SO(10)$
model, with no extra fields added. The argument for
this endeavor is that our approach is different in two significant ways
from previous analyses. First, we use a different method than
\cite{moha2,Dutta1} in fitting for the lepton--quark sector. Since this
fit is technically rather difficult, and moreover, since the results of this
fit define the parameter space in which one can search for an acceptable prediction for
the neutrino sector, we think that it is important to have an
alternative approach. Second, rather than relying on precomputed
values of quark sector parameters at GUT scale, we use as inputs $M_Z$ scale
values, and run them up to unification scale. This allows for more flexibility
and we think more reliable predictions for the parameter values at GUT scale.
With these modifications in our approach, we find that we agree with some
results obtained in \cite{moha2,Dutta1} (in particular, the fact that
type--II seesaw does not work well when the CKM phase is in the first quadrant),
but not with others. Most interesting, we find that it is possible to fit
the neutrino sector in the minimal $SO(10)$ model,
in the case when type--I seesaw contribution to neutrino mass dominates. We also present
a mixed scenario which gives excellent agreement with the neutrino data.


The  paper is organized as follows. In the next section we give a quick
overview of the features of the minimal $SO(10)$ model relevant for our
purpose. In section III we address the problem of fitting the lepton--quark
sector in this framework. We also define the experimentally allowed range
in which the input parameters (quark and lepton masses at $M_Z$ scale)
are allowed to vary.  We start section IV with a quick overview of
the phenomenological constraints on the neutrino sector. There we provide a
very good fit to all the fermion masses and mixings using type--I seesaw. We follow
by analyzing the predictions of the minimal $SO(10)$ model in the
case when type--II seesaw is the dominant contribution to neutrino masses.
We then analyze the predictions in a  type--I seesaw dominance scenario,
and in a scenario when both contributions (type--I and type--II) have
roughly the same magnitude. We end with our conclusions in Sec. V.

\section{The minimal SO(10) model}

The model we consider in this paper is an supersymmetric $SO(10)$
 model where the masses
of the fermions are given by coupling with only two Higgs
multiplets: a {\bf 10} and a $\overline{\bf 126}$ \cite{babu}.
Both   the ${\bf H_{10}}$ and  $\bf H_{\overline{126}}$ contain
Higgs multiplets which are {\bf (2,2)} under  the $SU(2)_L \times
SU(2)_R$ subgroup. Most of these {\bf (2,2)} Higgses acquire mass
at the GUT scale. However, one pair of Higgs doublets $H_u$ and
$H_d$ (which generally are linear combinations of the original
ones) will stay light. (Details about the Higgs multiplet decomposition and
$SO(10)$ breaking can be found, for example,  
in \cite{Bajc_sb,Fukuyama_sb,Aulakh_sb,nasri1}).
 Upon breaking of the $SU(2)_L \times
U(1)_Y$ symmetry of the Standard Model, the vacuum expectation
value of the $H_u$ doublet will give mass to the up-type quarks
and will generate a Dirac mass term for the neutrinos, while  the
vacuum expectation value of the $H_d$ doublet will give mass to
the down-type quarks and the charged leptons.

 The mass matrices for quarks and leptons wil then have the following
form:
\bea{mm}
M_u &  = & \kappa_u Y_{10} + \kappa'_u Y_{126} \nonumber \\
M_d &  = & \kappa_d Y_{10} + \kappa'_d Y_{126} \nonumber \\
M_{\nu}^D &  = & \kappa_u Y_{10} - 3 \kappa'_u Y_{126} \nonumber \\
M_l &  = & \kappa_d Y_{10} -3 \kappa'_d Y_{126}~  \eea where
$Y_{10}, Y_{126}$ are the Yukawa coefficients for the coupling of
the fermions to the ${\bf H_{10}}$ and  ${\bf H_{\overline{126}}}$
multiplets respectively. Note that in the above equations the
parameters $\kappa_{u,d}, \kappa'_{u,d}$ as well as the Yukawa
matrices are in general complex, thus insuring that the  fermion
mass matrices will contain CP violating phases.

The $SO(10)$ ${\bf H_{\overline{126}}}$ multiplet also contains a
$({\bf \overline{10}},{\bf 3},{\bf 1})$ and a $({\bf 10},{\bf 1},{\bf
3})$ Pati-Salam multiplets. The Higgs fields which are color
singlets and $SU(2)_R$/$SU(2)_L$ triplets (denoted by $\Delta_R$
and $\Delta_L$) may provide Majorana mass term for the
right--handed and the left--handed neutrinos. One then has:
\bea{mmaj}
M_{\nu R} & = & \left\langle \Delta_R\right\rangle Y_{126} \nonumber \\
M_{\nu L} & = & \left\langle \Delta_L \right \rangle Y_{126}~.
\eea
If the vacuum expectation of the $\Delta_R$ triplet is around $10^{14}$ GeV
then
the Majorana mass term for the right--handed neutrinos will give rise,
through the seesaw mechanism, to left--handed neutrino masses of order
 eV. On the other hand, the VEV of $\Del_L$
contributes directly to the left--handed neutrino mass matrix
(this contribution is called type--II seesaw), so this requires
that the $\left\langle \Delta_L \right\rangle$ is either zero or
at most of order eV. This requirement is 
satisfied naturally in such models, since $\Del_L$ generally
aquires a VEV of order 
$M_Z^2 /\langle \Del_R \rangle$ \cite{seesaw}.

\section{Lepton and quark masses and mixings}

Our first task is to account for the observed lepton and quark
masses, and for the measured values of the CKM matrix elements. By
expressing the Yukawa matrices $Y_{10}$ and $Y_{126}$ in Eqs.
(\ref{mm}) in favor of $M_u$ and $M_d$, we get a linear relation
between the lepton and quark mass matrices; at GUT scale:
\beq{ML_eq} M_l\ =\ a\ M_u + b\ M_d \ , \eeq where $a$ and $b$ are
a combinations of the $\kappa_{u,d}, \kappa'_{u,d}$ parameters in
Eq. (\ref{mm}). For simplicity let's work in a basis where $M_d$
is diagonal (this can be done without loss of generality). Then, $
M_u = {\cal V}_{CKM}^T M_u^d \ {\cal V}_{CKM}$ with $M_u^d$ the
diagonal up--quark mass matrix. If we allow the entries in the
diagonal quark mass matrices to be complex:
  $M_u^d = diag\{m_u e^{i a_u}, m_c e^{i a_c}, m_t e^{i a_t} \} $,
  $M_d^d = diag\{m_d e^{i b_d}, m_s e^{i b_s}, m_b e^{i b_b} \} $,
then the CKM matrix can be written in its standard form as a
function of three real angles and a phase:
\beq{ckm_def}
{\cal V}_{CKM} \ = \ \left(
\begin{array}{ccc}
c_{12} c_{13} & s_{12} c_{13} & s_{13} e^{-i \del} \\
-s_{12} c_{23} -c_{12} s_{23} s_{13} e^{i \del} &
   c_{12} c_{23} - s_{12} s_{23} s_{13} e^{i \del} & s_{23} c_{13} \\
s_{12} s_{23} -c_{12} c_{23} s_{13} e^{i \del} &
   -c_{12} s_{23} - s_{12} c_{23} s_{13} e^{i \del} & c_{23}
   c_{13}
\end{array} \right)~.
\eeq Since their phases can be absorbed in the definitions of the
$a_i, b_i$ parameters, we will take the coefficients $a, b$ to be
real, too.  One of the quark mass phases can be set to zero
without loss of generality, we set $b_b =0$.  It should be noted
that a common phase of $\{a,~b\}$, which we denote as $\sigma$,
will appear in the Dirac and Majorana mass matrices of the
neutrinos, and will be relevant to the study of neutrino
oscillations.

The  relation (\ref{ML_eq}) will generally impose some constraints
on the masses of the quarks and leptons.  For example, if we take
all the phases  to be zero (or $\pi$), then on the right-hand side
of the equation there are just two unknowns, the coefficients $a,
b$. On the other hand, the eigenvalues of the lepton mass matrix
are known, which will give us 3 equations. It is not obvious,
then, that this system can be solved; however, early analysis
\cite{lavoura,babu}
 shows that solutions exist, in the range of experimentally
allowed values, provided that the quark masses satisfy some constraints.

Newer studies \cite{fukuyama,moha1,moha2} 
allow for (some) phases
to be non-zero, and thus relaxes somewhat the constraints on quark
masses. However, it is interesting to note that these solutions
are not very different from the  purely real case. That is, most
of the phases involved have to be close to zero (or $\pi$), and
the values of the parameters do not change by much. We shall
explain this in the following.

The algebraic problem of solving for the lepton masses in the case
when the elements of the matrices $M_u, M_d$ are complex is quite
difficult. This would involve solving a system of 3 polynomial
equations of degree six in unknown quantities $a, b$. Most of the
analysis so far is done by numerical simulations (some analytical
results are obtained for the case of 2nd and 3rd families only
\cite{bajc,Bajc2}). In this section we attempt to solve the full
problem (with all the phases nonzero) in a semi-analytical manner,
that is, by identifying the dominant terms in the equations and
obtaining an approximate solution in the first step, which can
then be made more accurate by successive iterations.

 Due to the hierarchy between the eigenvalues of the
lepton mass matrix $m_e \ll \mm \ll \mtau$ one can suspect that
the mass matrix itself has a hierarchical form. This assumption is
supported by the observation that the off-diagonal elements of
$M_l$ are indeed hierarchical; for example  $L_{13}/ L_{23} \simeq
{\cal V}_{31} / {\cal V}_{31} \ll 1$. ($L_{ij}$ is a short--hand
notation for $(M_l)_{ij}$, ${\cal V}_{ij}$ are the $ij$ elements
of ${\cal V}_{CKM}$.) Then, the three equations for the invariants
of the real matrix $L L^{\dag}$ (the trace, the determinant and
the sum of its $2 \times 2$ determinants) become: \bea{Leq1}
| L_{33} |^2 + 2 | L_{23} |^2 & \simeq & \mtau^2 \nonumber \\
| L_{22} L_{33} - L_{23}^2|^2  & \simeq &  \mm^2 \mtau^2
 \nonumber \\
\hbox{Det} [ L L^\dag ] & = & m_e^2 \mm^2  \mtau^2 \ .
\eea

We find it is convenient to work in terms of the dimensionless
parameters $\tilde{m}_i = m_i/\mtau ~(i=e,\mu), \ta = a \tmt, \tb
= b \tmb$, and the ratios $r_c = m_c/m_t, r_s = m_s/m_b, r_u =
m_u/m_t, r_d =m_d/m_b$. Explicitly from the equations above in
terms of these parameters, we obtain: \bea{Leq3} \tilde{L}_{33} =
e^{i\al_1} & = &
 \ta {\cal V}_{33}^2 + \tb \ e^{-i z_3}
\nonumber \\
\tilde{\Del}_{23} = \tmm e^{i \al_2} & = &
(\tb \ r_s + \ta \ r_c e^{i z_2}) \tilde{L}_{33}
 + \ta \ \tb \ {\cal V}_{32}^2 e^{i (b_b -b_s)} \nonumber \\
\tilde{\Del} = \tme \tmm e^{i\al_3} &  = &
\tb \ r_d \ \tilde{\Del}_{23} - \ta^2 \ \tb \ e^{i (a_t - b_d)}
\biggl(  r_s ({\cal V}_{31} {\cal V}_{33})^2 \nonumber \\
& &
+ 2 r_c {\cal V}_{31} {\cal V}_{32} {\cal V}_{21} {\cal V}_{22} e^{i(z_2 -z_3)}
+ r_c {\cal V}_{31}^2 {\cal V}_{22} {\cal V}_{33} e^{i z_2} \biggr)
\eea
Here we have kept only the leading terms, using
 $r_u, r_d \ll r_c, r_s \ll 1$. Moreover,
 note that
only phase differences like $  a_i - b_i, a_i -a_j$
can be determinated from Eq. (\ref{ML_eq}); therefore,
by multiplying with overall phases,
we have written Eqs. (\ref{Leq3}) in terms of these differences
(with the notation $z_i = a_i - b_i$).

The key to solving this system is to recognize that there is some
tuning involved. Analyzing the first two equations leads to the
conclusion that $\ta, \tb \simeq {\cal O}(10)$. Then the phase
$z_3$ in the first equation should be close to $\pi$ so that the
two terms almost cancel each other. Similar cancellations happen
in the second and the third equations, which require respectively
that $b_b - b_s \simeq \pi$ and $b_d \simeq a_t$\footnote{Note
that taking these phase differences to $\pi$ or zero results in
exactly the mass signs which the analysis in \cite{fukuyama} found
to work for the real masses case.}. Also, in the third equation,
neglecting the small electron mass on the left hand side results
in: \beq{a_sol} \ta^2 \simeq { r_d \ \tmm \over r_s |{\cal V}_{31}
{\cal V}_{33}|^2 }~. \eeq For values of the parameters in the
experimentally feasible region, this is consistent with the above
estimate $\ta \simeq {\cal O}(10)$.

 Analytically solving Eqs. (\ref{Leq3}) with the approximations discussed
will provide  solutions
for the phases and parameters $\ta, \tb$
accurate to  the 10\% level. Using these first order results, one
can compute and put the neglected terms back in Eqs. (\ref{Leq1},\ref{Leq3}),
which can be solved again, thus defining an iterative procedure which can
be implemented numerically, and
brings us arbitrarily close to the exact solution. We find that 5 to 10
iterations are usually sufficient to recover the $\mu$ and $\tau$  masses with
better than 0.1\% accuracy ($m_e$ can be brought to a fixed value by
multiplying with an overall coefficient).

We end this section with some comments on the range of input parameters
(masses and phases) which allow for a solution to Eq. (\ref{ML_eq}). As we
discussed above, the phases are either close to $\pi$ or to zero. This
is required by the necessity to almost cancel two large terms
in the right-hand side of Eqs. (\ref{Leq3}). One can see that the larger
the absolute magnitude of these terms (for example
$|\tilde{a} {\cal V}_{33}^2|$ and
$\tilde{b}$ in the first equation), the more stringent are the constraints
on the phases. The opposite is also true; the smaller the $\ta$ and $\tb$
parameters, the more the phases can deviate from $\pi$, and generally the
easier it is to solve the system. This means that lower values of
$\ta$,  $\tb$ are preferred; from Eq. (\ref{a_sol}), this implies
a preference for low values of the ratio $m_d/m_s$ \footnote{This also
means higher values for $| {\cal V}_{31}|$ are
preferred. Since
${\cal V}_{31} = s_{12} s_{23} -c_{12} s_{23} s_{13} e^{i \del}$, this
implies a preference for values of the CKM
phase $\del$ close to $\pi$ (as noted in \cite{moha2}).} (there is not much
scope to vary $\tmm$). It turns out that low values of $m_s$ and
large values of $m_c$ can also help, since they lower the absolute
magnitude of the larger term on the right-hand side of the equation
for   $\tilde{\Del}_{23}$ in (\ref{Leq3}). Previous analysis found indeed that
fitting for the lepton masses require a low value for $m_s$ \cite{Dutta1}.

\subsection{Low scale values and RGE running}

As was discussed in the above section, the relation (\ref{ML_eq})
implies some constraints on the quark masses (the lepton masses being
taken as input). That is, not all values
of quark masses consistent with the experimental results are
also consistent with the model we use.
Our purpose first is to identify these points in the parameter space defined
by the experimentally alowed values for quark masses,

 Let us then define what this parameter space is. Altought
the relations in the previous section hold at GUT scale, one must necessarily
start with the low energy values for our parameters.
We choose to use as input the values
of the  quark masses and the CKM angles at the $M_Z$ scale.
 Estimates of these quantities can be found
for example in \cite{koide}. However, we consider some of their numbers rather
too precise (for example, their error in estimating the masses of the
$s$ and $c$ quarks are only 25\%, respectively 15\%, while the
corresponding errors in PDG \cite{pdg} are much larger).
 Therefore, in the interest
of making the parameter space as large as possible, we use the following
values:
\begin{itemize}
\item for the second family: 70 MeV $< m_s(M_Z) < $ 95 MeV
\footnote{Note that the lower limit for $m_s(M_Z)$ is rather low
compared with \cite{koide}; however, the value at 2 GeV scale is
well within the limits cited in \cite{pdg}. Lattice results also
seem to favor smaller values of $m_s$(2 GeV) \cite{hashimoto}.};
650 MeV $< m_c(M_Z) <$ 850 MeV. With a running factor from $M_Z$
to 2 GeV of around 1.7, these limits would translate to values at
2 GeV scale of: 120 MeV $\lesssim m_s(\hbox{2 GeV}) \lesssim $ 160
MeV ; 1.1 GeV $\lesssim m_c(\hbox{2 GeV}) \lesssim $ 1.44 GeV.
Lattice estimations $(m_c/m_s)_{2 GeV} \simeq 12$  \cite{Gupta}
would indicate a value in the lower part of the range for $m_s$,
and a upper part for $m_c$. \item for the light quarks: here
generally the ratio of quark masses are more trustworthy than
limits on the masses themselves; we therefore use use  $17 <
m_s/m_d < 23 $ (as noted in the previous section, high values of
this ratio are preferred), and $ 0.3 < m_u/m_d < 0.7$. We  note
here that $m_u$  is a parameter which does not affect the results
much. \item for the heavy quarks: 2.9 GeV $ < m_b(M_Z) < $ 3.11
GeV, (or  4.23 GeV $ < m_b(m_b) < $ 4.54 GeV)  and for the pole
top mass 171 GeV $ < M_t < $ 181 GeV (the corresponding $\bar{MS}$
mass is evaluated using the three loops relation, and comes out
about 10 GeV smaller). \item the CKM angles at $M_Z$ scale:
$$s_{12} = 0.222 \pm 0.003\ ,\ s_{23} = 0.04 \pm 0.004 \ ,
\ s_{13} = 0.0035 \pm 0.0015 . $$
\end{itemize}

For the gauge coupling constants we take  the following values at $M_Z$
scale: $\alpha_1 (M_Z) = 1/58.97, \alpha_2 (M_Z) = 1/29.61,
\alpha_3 (M_Z) = 0.118$. With these values at low scale
one can get unification
of coupling constants at the scale $M_{GUT} \sim 10^{16}$.
The exact value of $M_{GUT}$, as well as
the values of the fermions Yukawas at the unification scale,
will depend also on the supersymmetry breaking scale
($M_{SUSY}$) and $\tan \beta$, the 
ratio between the up-type and down-type SUSY Higgs
VEVs. We generally consider values of $M_{SUSY}$ between
200 GeV and 1 TeV, and $\tan \beta$ between 5 and 60.

Having chosen specific values of the parameters described above, we then
run the fermion Yukawa coupling and the quark sector mixing angles, first
from $M_Z$ to $M_{SUSY}$, using two-loop Standard Model renormalization
group equations; then we run from SUSY scale to the GUT scale using 
two loop\footnote{More precisely, we use the two-loop RGEs 
for the running of the gauge
coupling constants and the third family fermions ($b,t$ and $\tau$). To
evaluate the light fermion masses, we use the one-loop equations for
the ratios $m_{s,d}/m_b, m_{c,u}/m_t$ and $m_{\mu,e}/m_{\tau}$.
 This approximation is
justified, since the leading two-loop effect on the fermion masses comes
from the change in the values of gauge coupling constants at two-loop;
however, the contributions due to the gauge terms 
are family-independent and will not affect these ratios.}
SUSY RGEs \cite{barger}. After computing the neutrino mass matrix at GUT scale, we
run its elements back to $Z$ scale \cite{babuleung,Chankowski}
and evaluate the resulting
masses and mixing angles.

\section{Neutrino masses and mixings}

In the present framework, there are two contributions to neutrino masses.
First one has the canonical seesaw term:
\beq{t1_sw}
 (M_{\nu})_{seesawI} = M_\nu^D M_R^{-1} M_\nu^D \eeq
with $M_R = v_R Y_{126}$ and $M_\nu^D$ given by (\ref{mm}).
However, the existence in this model of the ($\bar{{\bf 10}}$,{\bf
3},{\bf 1}) Higgs multiplet implies the possibility of a direct
left-handed neutrino mass term when the $SU(2)_L$ Higgs triplet
$\Delta_L$ from this acquires a VEV $v_L$ (as it generally can be
expected to happen). The neutrino mass contribution of such a term
would be \beq{t2_sw}
 (M_{\nu})_{seesawII} = v_L Y_{126} = \lambda M_R \eeq
where $v_L = \gamma v_{weak}^2/v_R$ and $\gamma$ is a factor
depending on the specific form of the Higgs potential \cite{seesaw}.

The scale of the canonical seesaw contribution Eq. (\ref{t1_sw})
(which we call type--I seesaw in the following) to the left handed
neutrino mass matrix is given by $v_{weak}^2/v_R$. The
contribution of the type--II seesaw factor (Eq. (\ref{t2_sw})) is
of order $ \gamma v_{weak}^2/v_R$. One cannot know apriori how the
factor $\gamma$ compares with unity, therefore one cannot say
which type of seesaw dominates (or if they are of the same order
of magnitude). Therefore, in the following each case will be
analyzed  separately.

However, let us first review the current experimental data on the
neutrino mixing angles and mass splittings. Latest analysis
\cite{Maltoni} sets the following $3\sigma$ bounds:
\begin{itemize}
\item from $\nu_{\mu} - \nu_{\tau}$ oscillations:
$$ 1.4 \times 10^{-3}\ \hbox{eV}^2 \leq \Delta m_{23}^2 \leq
3.3 \times 10^{-3}\ \hbox{eV}^2 \ \ ;
\ \ 0.34 \leq \sin^2\theta_{23} \leq 0.66 \ ;  $$
with the best fit for $\Delta m_{23}^2 = 2.2 \times 10^{-3}\ \hbox{eV}^2$
and $\sin^2\theta_{23} = 0.5 $ (from atmospheric and K2K data).
\item from $\nu_{e} - \nu_{\mu}$ oscillations:
$$ 7.3 \times 10^{-5}\ \hbox{eV}^2 \leq \Delta m_{12}^2 \leq
9.1 \times 10^{-5}\ \hbox{eV}^2 \ \ ; \ \ 0.23 \leq
\sin^2\theta_{12} \leq 0.37 \ ;  $$ with the best fit for $\Delta
m_{12}^2 = 8.1 \times 10^{-5}\ \hbox{eV}^2$ and $\sin^2\theta_{12}
= 0.29 $ (from solar and KamLAND data). Note also that a
previously acceptable region with a somewhat higher mass splitting
$\Delta m_{12}^2 \simeq 1.4 \times 10^{-4}\ \hbox{eV}^2$ (the LMA
II solution \cite{fogli}) is excluded now at about $4\sigma$ by
the latest KamLAND data. \item finally, by using direct
constraints from the CHOOZ reactor experiment as well as combined
three-neutrino fitting of the atmospheric and solar oscillations,
one can set the following upper limit on the $\theta_{13}$ mixing
angle:
$$ \sin^2\theta_{13} \leq 0.022 \ .$$
\end{itemize}

The procedure we use in searching for a fit to neutrino sector
parameters is as follows. First the low scale values  of the
quark and lepton masses and the CKM matrix angles and phase are chosen.
(Generally we take a fixed value for $m_b$ and $m_t$, while the other
parameters are chosen randomly from a predefined range; however,
$m_b$ and $m_t$ can also be chosen randomly).
Next we pick a value for $\tan \beta$ and $M_{SUSY}$, and compute
the quark-lepton sector quantities at GUT scale. Here we determine
the relation between the lepton Yukawa couplings and quark Yukawa couplings,
which amounts to determining the parameters $a,b$ and phases $a_i,b_i$ in
Eq. (\ref{ML_eq}). The phases combinations $z_i = a_i + b_i$ are chosen
as input (that is, they are picked randomly), while $a,b$, and the
remaining two phases are obtained by the procedure of fitting
the lepton eigenvalues described in  Section III. Finally, we scan over
the parameters which appear in the neutrino sector (if the
neutrino mass matrix is either of type--I seesaw or of type--II seesaw, there
is only one phase $\sigma$;
if both types appear, there will be two extra parameters,
the relative magnitude and phase of the two contributions).

The rest of this section is devoted to
a detailed analysis of the predictions of the $SO(10)$ minimal model
for the neutrino
sector, in type--I, type--II and mixed scenarios. (Due to its relative simplicity,
we will start with the type--II case). However,
let us first summarize our results.
We find that in the type--II scenario, there is no good fit to the neutrino
sector if the CKM phase is consistent with experimental measurements
(around 60 deg). This is in agreement
with previous analysis \cite{moha2,Dutta1}; however,  our results are a bit
more encouraging, in that that for $\delta_{CKM} = \pi/2$ we find reasonably
good fits, which improve significantly with not very large increases in
the CKM phase. We can obtain marginal fits for $\delta_{CKM} $ as low as 80 deg.
 More interesting are the results for the type--I case; here
we can find good fits to the neutrino sector for values of $\delta_{CKM}$
as low as $50^o$, certainly consistent with experimental limits. As such
fits have been not found before, one might consider this to be the main result
of our paper. Also, we find that in the mixed case, there is possible
to obtain a good neutrino sector fit in the case when the contributions
coming from type--I and type--II are roughly equal in magnitude and of opposite
phase.
\subsection{Example of Type--I Seesaw Fit}

We give here a representative example of a fit obtained in a
type--I dominant case. This is obtained for $m_s(M_Z) = 0.07$ GeV,
$m_c(M_Z) = 0.85$ GeV, $m_b(M_Z) = 3$ GeV, $M_t =$ 174 GeV, $\tan
\beta = 40$ and $M_{SUSY} = 500$ GeV. The values of the quark and
lepton masses at GUT scale (in GeV) and the CKM angles are:

\bea{num}
\begin{array}{rlrlrl}
m_u &= 0.0006745 \; &  m_c & =0.3308  & m_t & =97.335  \\
m_d & =0.0009726 \; &  m_s & =0.02167   &  m_b & =1.1475   \\
m_e & =0.000344 \; & m_\mu & =0.0726 \ \  & m_\tau & =1.350 \\
s_{12} & = 0.2248 &  s_{23} & = 0.03278 & s_{13} & = 0.00216 \\
& &  \delta_{CKM} & = 1.193 \ . & &
\end{array}
\eea

Here the masses are defined as $m_u = Y_u \sin\beta \ v_0$, $m_d =
Y_d \cos\beta \ v_0$ where $Y_u, Y_d$ are the corresponding Yukawa
couplings, and $v_0 = 174$ GeV is the SM Higgs vacuum expectation
value\footnote{ One can write Eqs. (\ref{ML_eq}), (\ref{nu_mpp})
in terms of either the Yukawa couplings of the leptons and quarks,
or their masses (that is, Yukawa couplings times running Higgs
VEVs). In this paper we use the Yukawa couplings, but we multiply
by the Higgs VEVs at the SUSY scale for simplicity of
presentation. One can easily check then when going from one
convention to the other, just the parameter $a$ rescales, while
$b$ does  not change.}. The values of GUT scale phases (in
radians) and $a,b$ parameters are given by: \bea{num2}
\begin{array}{rlrlrl}
 a_u & = 0.881  &    a_c & = 0.32678 \ &  a_t & = 3.0382  \\
b_d & = 3.63235 &     b_s & = 3.23784 &  b_b & = 0.  \\
  a & = 0.08136 \ &  b & = 5.9797 &  \sigma & = 3.244 \ .
 \end{array}
\eea
With these inputs, one can evaluate all mass matrices at GUT scale. In order
to compute the neutrino mass matrix at $M_Z$ scale, we use the running factors
$r_{22} = 1.06, r_{23} = 1.03$, where
$$ r_{22} = \left( { M_{\nu ij} \over M_{\nu 33}} \right)_{M_Z}
\slash \left( { M_{\nu ij} \over M_{\nu 33}} \right)_{M_{GUT}} \ , \ \
r_{23} = \left( { M_{\nu i3} \over M_{\nu 33}} \right)_{M_Z}
\slash \left( { M_{\nu i3} \over M_{\nu 33}} \right)_{M_{GUT}} \ ,
$$
with $ i,j = 1,2$. The elements of the neutrino matrix above
are evaluated in a basis where the lepton mass matrix is diagonal.

One then obtains for the neutrino parameters at low scale:
$$\Delta m^2_{23}/\Delta m^2_{12} \simeq 24 \ , \
\sin ^2 \theta_{12}  \simeq 0.27 \ , \ \sin ^2 2\theta_{23} \simeq 0.90\
, \ \sin ^2 2\theta_{13} \simeq 0.08 \ .$$
Note here that only the atmospheric
angle is close to the experimental limit, the
solar angle and the mass spliting ratio being close to the preferred values.
The elements of the diagonal
neutrino mass matrix are
$$ m_{\nu i} \ \simeq \ \{0.0021 \exp(0.11i)\ ,
\ 0.0098\exp(-3.06i)\ ,\ 0.048  \}
$$
in eV,
with a normalization $\Delta m_{23}^2 = 2.2 \times 10^{-3}$eV$^2$.
The phases of the first two masses are the Majorana phases (in radians).
Moreover,
the Dirac phase appearing in the MNS matrix is $\phi_D = -0.23$rad,
and one evaluates the effective neutrino mass for the neutrinoless double
beta decay process to be
$$ | \sum U_{e i}^2 m_{\nu i}| \ \simeq \ 0.009 \ \hbox{eV} \ .
$$

\subsection{Type--II seesaw}

Much of the recent work on the neutrino sector in the minimal
$SO(10)$ has concentrated on the scenario when the type--II seesaw
contribution to neutrino masses is dominant. The reason for the
interest in this case is that, with:
$$M_{\nu} \ \sim \  M_R  \ \sim \ M_l - M_d~. $$
$b-\tau$ unification at the GUT scale, $m_b \simeq m_{\tau}$,
 naturally leads to a small value of
$(M_{\nu})_{33}$ and hence large mixing in the 2-3 sector
\cite{bajc}. However, while the general argument holds, it has
been difficult (or impossible) to fit both large
$\theta_{\hbox{atm}}$ and the hierarchy between the solar and
atmospheric mass splittings at once. In this section we will try
to show why this is so, and under which conditions this might be
achievable.

We will use the same conventions as in section III (that is, we work
in a basis where $M_d$ is diagonal, and the parameters $a, b$ and
$(M_d)_{33} = m_b$ are real and positive). However, in the construction
of the neutrino mass matrices there will be an extra phase besides
those which were relevant for the quark-lepton mass matrices. This phase
$\sigma$ can be though as an overall phase of $M_l$. One then has:
\bea{nu_mpp}
M_R & = & y ( e^{i \sig}  {M_l} - {M_d}  ) \nonumber \\
a M_\nu^D & = & -(b e^{i \sig} + 2) {M_l} e^{i \sig} + 3 {M_d} \ .
\eea

Following the analysis in Sec. III one can write: \bea{le_mm}
({M_l})_{22} & \simeq & | b | \ m_s \ e^{i b_2} \nonumber \\
({M_l})_{23} & \simeq & a \ m_t \ e^{i a_3}
         {\cal V}_{32} {\cal V}_{33} \nonumber \\
({M_l})_{33} & = & a \ m_t  \ e^{i a_3} + b \ m_b\
          \simeq \ m_{\tau} e^{i \al} \ ,
\eea
with $a_t$ close to $\pi$ and $\alpha + b_s = \epsilon$ close to zero.
Then, the neutrino mass matrix will be proportional to:
\beq{nu_mm1}
(M_{\nu})_{[2,3]} \sim {M_l} - {M_d}^d e^{-i \sig}
\sim \left(
\begin{array}{cc}
m_s e^{i (\epsilon - \alpha)} (b - e^{-i\sig}) & m_{23} \\
m_{23} & m_{\tau} e^{i \alpha} - m_b e^{-i \sig}
\end{array} \right)
\eeq
Note also that $m_{23} = \ m_t \ e^{i a_t} {\cal V}_{32} {\cal V}_{33}$
is almost real positive, and due to the fact that $a m_t \simeq b m_b$
and $m_b {\cal V}_{32} \simeq m_s$, the 22 and 23 elements in the
neutrino mass matrix are roughly of the same order of magnitude (in
practice, one get $m_{23}$ somewhat larger than $m_{22}$). One then
sees that if the phase $\sigma $ is chosen such that the two terms
in the 33 mass matrix element cancel each other (that is
$\sigma \simeq - \alpha$) then there will be large mixing in the 2-3
sector, with:
$$ \tan (\theta_{\nu})_{23} \simeq \left| {m_{23} \over m_{33} } \right|. $$

However, this is not the whole story. One needs also some hierarchy between
the atmospheric and solar neutrino mass splittings:
$$ {\Delta m^2_{sol} \over \Delta m^2_{atm}} \  = \ r
\ \lesssim \ {1 \over 20 } $$
(based on the experimental measurements of neutrino parameters reviewed
in the previous section). In terms of the eigenvalues $m_2, m_3$
 of the mass matrix (\ref{nu_mm1}), one then has:
\beq{mass_eq}
 \ {m_2^2 \over m_3^2} \ \simeq \left( {|m_{22} m_{33} - m_{23}^2| \over
 |m_{22}^2| + |m_{33}^2| + 2 |m_{23}|^2 } \right)^2
\ \lesssim \ {1 \over 20 } \ .\eeq In order for this to hold, one
needs a cancellation between the $m_{22} m_{33}$ and $m_{23}^2$
terms in the numerator of the above fraction. This in turn imposes
a constraint on the phases involved: \beq{phase_eq} \phi =
\hbox{Arg}(m_{\tau} - m_b e^{- i (\sig -\al)}) \simeq 0 \ . \eeq

More detailed analysis shows that it is not possible (or very
difficult) to get $\tan
(\theta_{\nu})_{23}$ larger than 1 while satisfying the relation
(\ref{mass_eq}) between eigenvalues. However, this  will create problems with
the atmospheric mixing angle. The PMNS matrix is
$$ U_{PMNS} = U_l^\dagger U_{\nu}$$
where $U_l, U_{\nu}$ are the matrices which diagonalize the lepton
and neutrino mass matrices, respectively. Since the lepton mass
matrix has a hierarchical form, the $U_l$ matrix is close to
unity, with $(U_l)_{i3} \simeq (0,x,1)$, where $ x =  (
m_{23}/m_{33})_l^*$. The atmospheric mixing angle will then be:
$$ \tan \theta_{\hbox{atm}} \ \simeq \ \left|
\frac{ \left( { m_{23} \over m_{33} } \right)_{\nu}^*
-  \left( { m_{23} \over m_{33} }\right)_{l}^*}{
 1 + \left( { m_{23} \over m_{33} } \right)_{\nu}^*
 \left( { m_{23} \over m_{33} }\right)_{l}}
\right| $$
 where the $l$ and $\nu$ lower indices make clear that we are discussing
elements of the lepton and neutrino mass matrices.
Note, however, that  Eq. (\ref{phase_eq}) implies that
$(m_{33})_l$ and $(m_{33})_{\nu}$ have the same phase of $m_\tau$;
then, since $(m_{23})_l = (m_{23})_{\nu}$, the net effect of the
rotation coming from the lepton sector is to reduce the 2-3 mixing
angle. Practically, since $|x| \approx 0.2 $, even if one has  a
value of $\tan  (\theta_{\nu})_{23}$ close to one from the
neutrino mass matrix,
 $\tan \theta_{\hbox{atm}}$ will become of order 0.7 after rotation in the
lepton sector is taken into account.

\begin{figure}[t!] 
\centerline{
   \includegraphics[height=2.in]{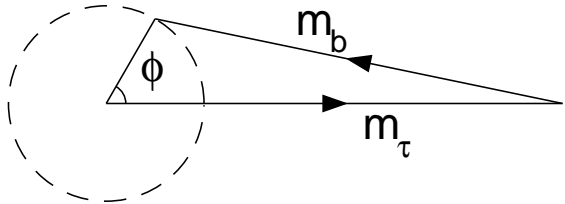}
   }
\caption{Graphical representation of the relationship between the $b$ and
$\tau$ masses at the GUT scale and the magnitude and phase  of
the 33 element of the neutrino mass matrix.}
\label{fig_1}
\end{figure}

This situation is represented graphically in Fig. \ref{fig_1}. As
discussed above, $\phi =0$ corresponds to the case most favorable
for getting the right solar-atmospheric mass splitting ratio,
while $\phi = \pi$ corresponds to the case of maximal mixing
angle. In practice this means that most reasonable fits are
actually obtained when the angle $\phi$ is close to around $\pi/2$
(otherwise generally either the angle or the mass ratio are too
small) \footnote{Contributions from the phases in the lepton mass
matrix can also improve the goodness of the fit (for example, if
$\epsilon$ is significantly different from zero, or $\alpha$
different from $\pi$). However, this generally requires that the
parameters $\ta, \tb$ have low values (as explained in section
III). Hence we see that neutrino sector also prefers
$\delta_{CKM}$ in the second quadrant and low $m_s$.}. Note
however that  $\phi = \pi$ (or any value greater than about
$\pi/2$) would require that at GUT scale $m_b > m_{\tau}$. We may
infer that in order to obtain a large mixing in the 2-3 sector one
needs that at GUT scale $m_b$ should be close to $m_{\tau}$.

\begin{figure}[t!] 
\centerline{
   \includegraphics[height=3.in]{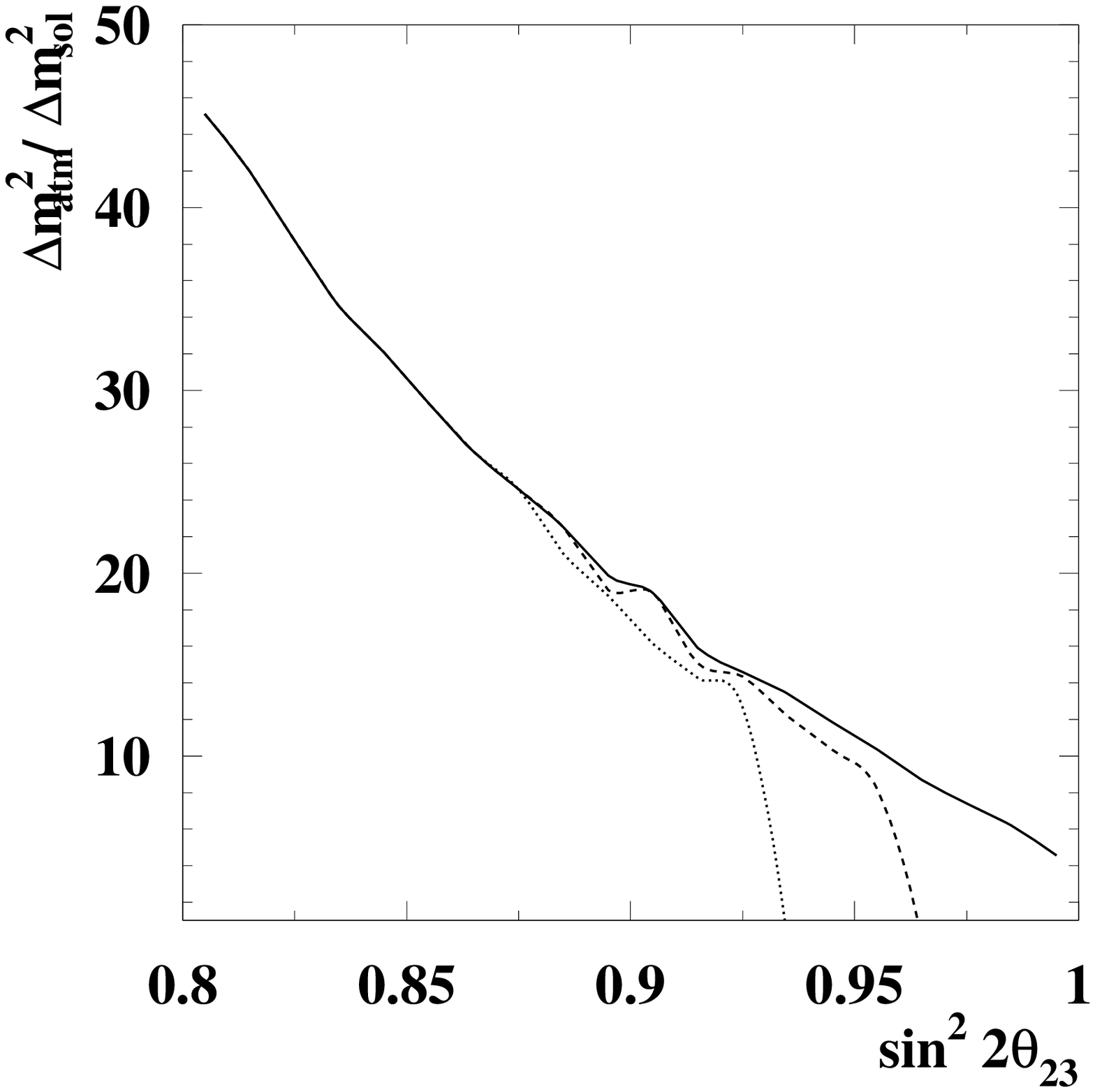}
   \includegraphics[height=3.in]{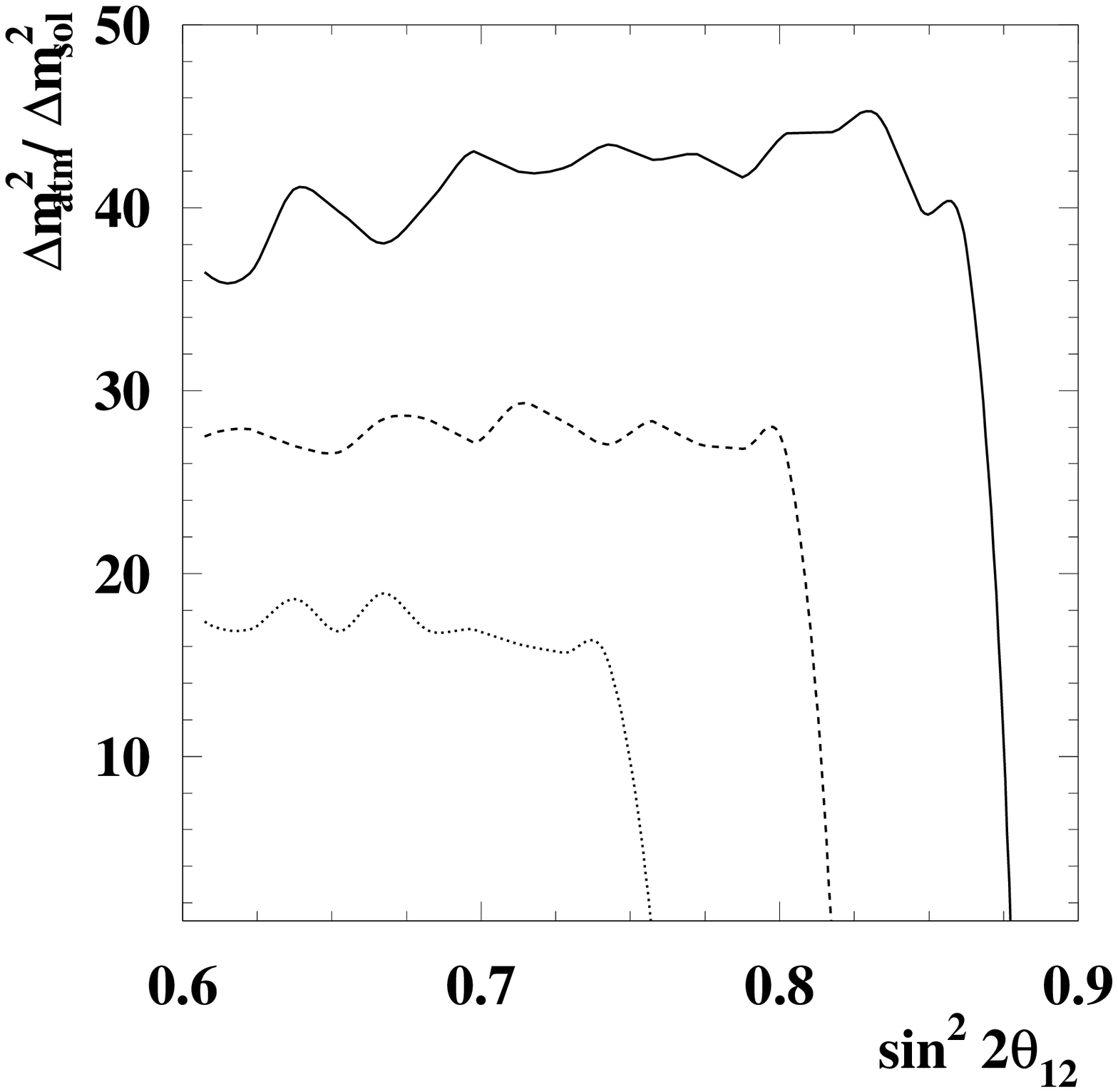}
   }
\caption{Left: maximum of the atmospheric/solar mass splitting ratio
$R_{a/s}$ as a function of the
atmospheric mixing angle $\theta_{23}$, with cuts on the solar angle
$\sin^2 2\theta_{12} > 0.7$ (dotted line),  $ 0.65$ (dashed)
and no cut (solid).
 Right: maximum of $R_{a/s}$ as a function of the
solar mixing angle $\theta_{12}$, with cuts on the atmospheric angle
$\sin^2 2\theta_{23} > 0.9$ (dotted line),  $ 0.85$ (dashed)
and $ 0.8$ (solid). }
\label{fig_90_18}
\end{figure}

This in turn can be insured by requiring large $m_b(M_Z)$  and/or
$\tan \beta$. For example, in Fig. \ref{fig_90_18} we present the results obtained for values
$m_b(M_Z) = 3.11$ GeV, $m_t = 181$GeV, $M_{SUSY} = 500$GeV and
$\tan \beta = 50$ (with these values, the ratio $r_{b/\tau} = m_b / m_{\tau}
(M_{GUT}) \simeq 0.96$). Also, here we set the CKM phase $\delta = 90\deg$,
and let the other quark sector parameters  vary between  the limits
discussed in section IIIA.
 The left panel shows the maximum
atmospheric/solar mass splitting ratio $R_{a/s}$
as a function of the atmosperic mixing angle $\theta_{atm} = \theta_{23}$.
 The three different lines correspond to different cuts on the
solar mixing angle: $\sin^2 2\theta_{12} > 0.7$ (dotted),
$\sin^2 2\theta_{12} > 0.65$ (dashed) and no cut (solid). One can
observe here the correlation betewee large atmospheric mixing and small
atmospheric-solar mass ratio.

Conversely, the right panel
shows the maximum of the ratio $R_{a/s}$ as a function of the
solar mixing angle $\theta_{sol} = \theta_{12}$. The three different lines
correspond to cuts on the atmospheric mixing angle:
$\sin^2 2\theta_{23} > 0.9$ (dotted),  $\sin^2 2\theta_{23} > 0.85$ (dashed)
and $\sin^2 2\theta_{23} > 0.8$ (solid line). We note here that the
correlation between the solar angle and the mass ratio has the form of a step
function (abrupt decrease in $R_{a/s}$ once $\sin^2 2\theta_{12}$ goes
over a certain threshold), while there seems to be a close to linear
correlation between the maximal solar and atmospheric angles.

It is interesting to consider how these results change if the paramenters
 $m_b(M_Z),m_t, M_{SUSY}$ and/or $\tan \beta$
are modified. One  finds out that the neutrino sector results have
a strong  dependence on the parameter $\tan \beta$. For example, if one
keeps the parameters used in Fig. \ref{fig_90_18} fixed but increases
$\tan \beta$, one finds that the fit for the atmospheric angle -
atmospheric/solar mass ratio improves to a certain amount. That can be traced
to the fact that the ratio $r_{b/\tau}$ increases with $\tan \beta$.
However, one also finds that the solar angle generally gets smaller.
This happens because there is a correlation between the solar angle
and the value of the $s$ quark mass at GUT scale; namely $\theta_{12}$
increases with $r_{s/b} = m_s/m_b(M_{GUT})$. On the other hand,
the ratio $r_{s/b}$ decreases with increasing $\tan \beta$.

\begin{figure}[t!] 
\centerline{
   \includegraphics[height=3.in]{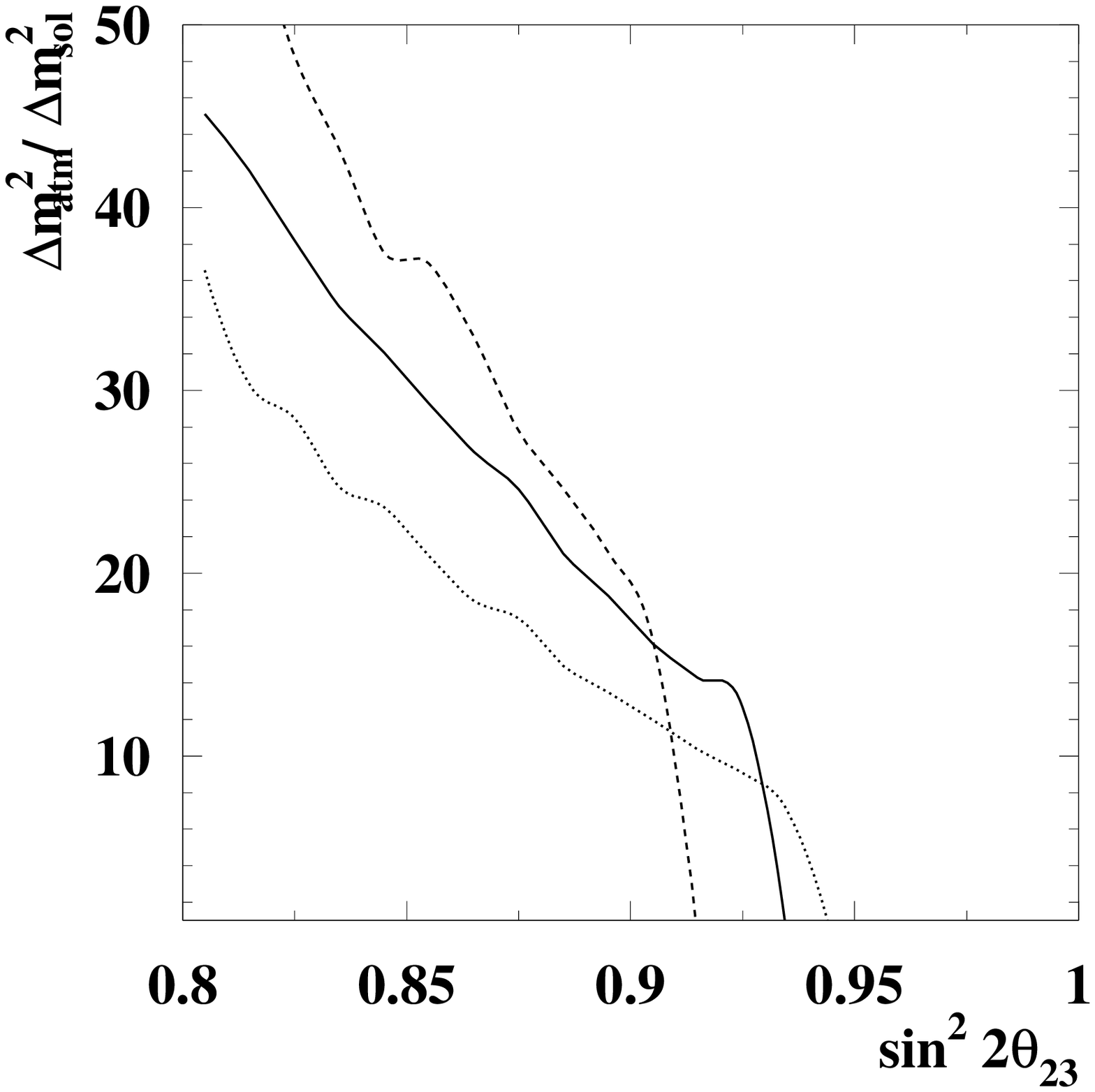}
   \includegraphics[height=3.in]{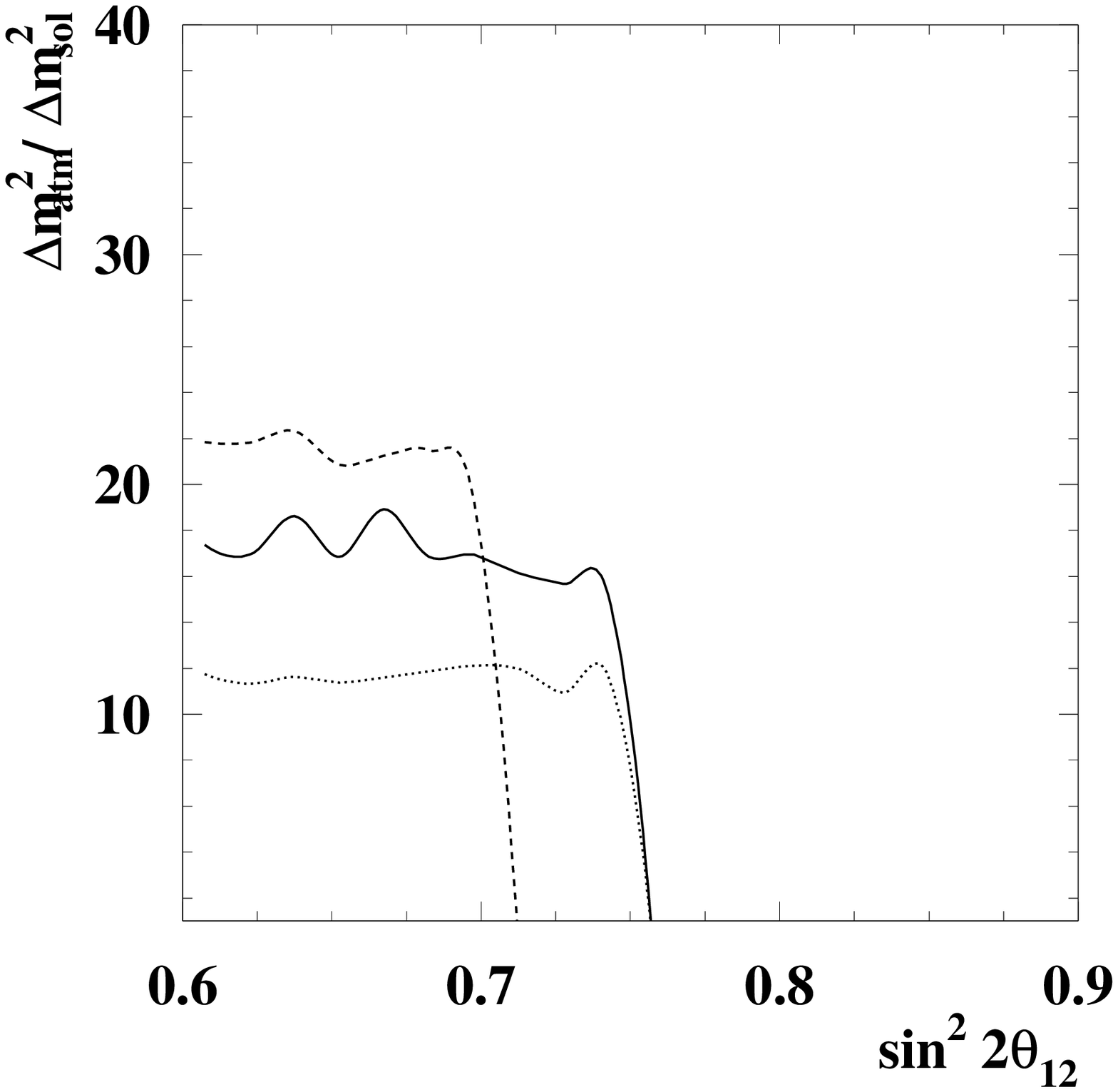}
   }
\caption{Maximum of the ratio $R_{a/s}$ as a function of the
atmospheric mixing angle $\theta_{23}$ (left)
and solar angle (right), for $\tan \beta = 40$ (dotted line),
50 (solid) and 55 (dashed). Additional cuts are
$\sin^2 2\theta_{12} > 0.7$ (left) and $\sin^2 2\theta_{23} > 0.9$ (right).}
\label{fig_23comb}
\end{figure}

Fig. \ref{fig_23comb} exemplifies this behaviour. The three lines
correspond to the maximum value for the mass splitting ratio $R_{a/s}$,
at values of  $\tan \beta = 40$ (dotted),  $\tan \beta = 50$ (solid)
and $\tan \beta = 55$ (dashed line). A cut on the solar angle
$\sin^2 2\theta_{12} > 0.7$ is also imposed in the left panel,
and a cut on the atmospheric angle $\sin^2 2\theta_{12} > 0.9$
in the right panel. One can see that at larger
values for $\tan \beta$ one might potentially get better fits for
atmospheric angle and the atmospheric/solar mass ratio;
however, the constraint
on the solar angle becomes more restrictive.

Smaller variations of the neutrino sector results will follow modification
of the parameters  $m_b(M_Z),m_t$ and $M_{SUSY}$. However, these variations
follow the same pattern as above: that is, an improvement in the fit
for the atmospheric angle due to the increase of the ratio $r_{b/\tau}$
(which can be due to an increase in $m_t$, or a decrease in $M_{SUSY}$)
coincide with a worsening of the fit for the solar angle. As a consequence,
the results presented in Figs. \ref{fig_90_18}, \ref{fig_23comb} can
be improved only marginally. Scanning over a range of parameter space
$2.9 \hbox{GeV} < m_b(M_Z) < 3.11 \hbox{GeV}, \
174 \hbox{GeV} < m_t < 181 \hbox{GeV}, \
500 \hbox{GeV} < M_{SUSY} < 1 \hbox{TeV}$ and $10 < \tan \beta < 60$
\footnote{In practice we find that the best results are obtained
for large $m_b(M_Z)$, large $m_t$ and large $\tan \beta$ (such that
 $r_{b/\tau}$ is between 0.96 and 1).},
we find the best fit to the neutrino sector to be
$\sin^2 2\theta_{23} \simeq 0.88$, $\sin^2 2\theta_{12} \simeq 0.74$ and
the atmospheric/solar mass splitting ratio $R_{a/s} \simeq 24$. We note that
although these numbers provide a somewhat marginal
fit to the experimental results
(the mixing angles are close to the exclusion limit, while the value for
mass ratio is central)
they are still allowed.

\begin{figure}[t!] 
\centerline{
   \includegraphics[height=3.in]{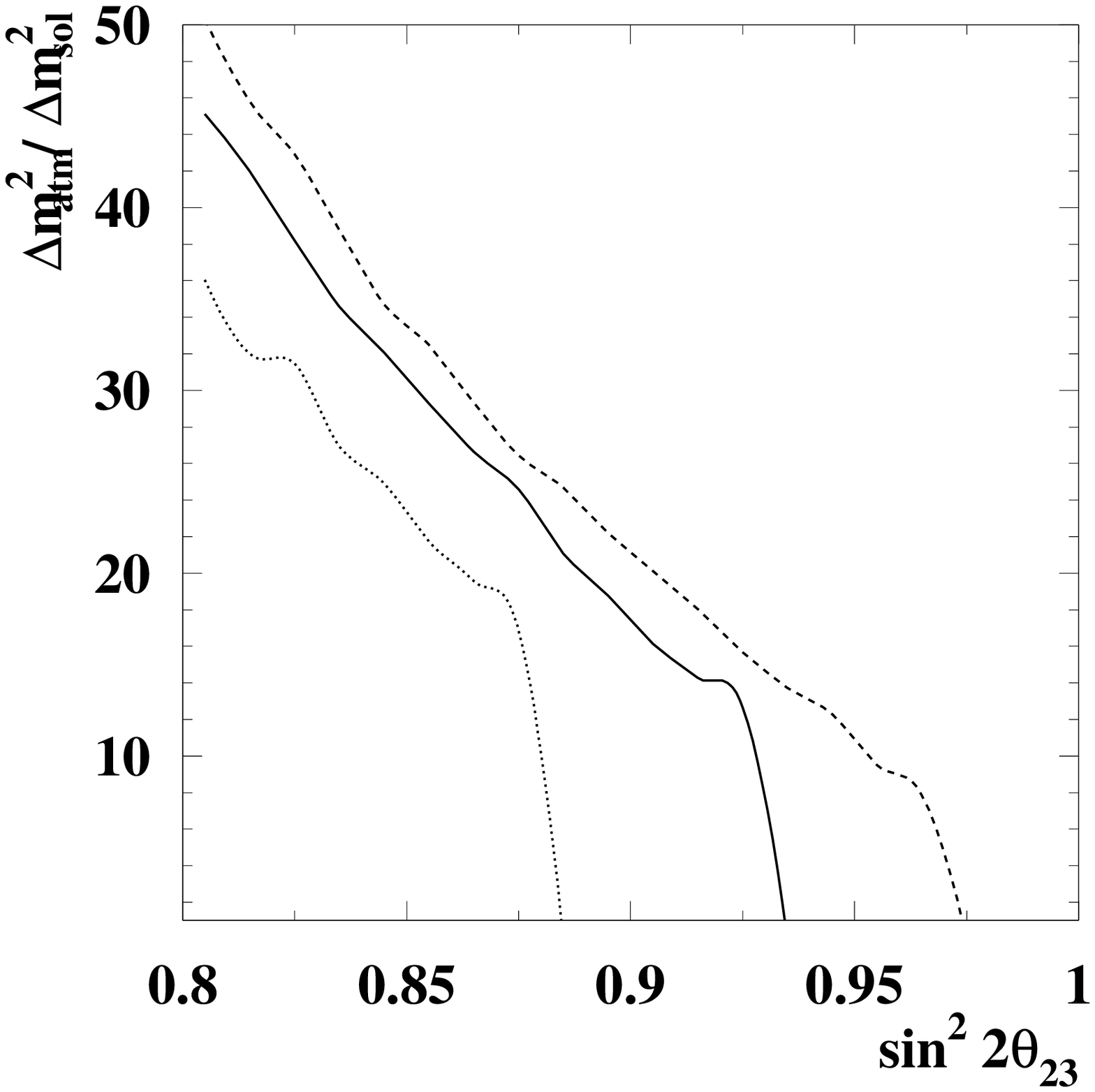}
   \includegraphics[height=3.in]{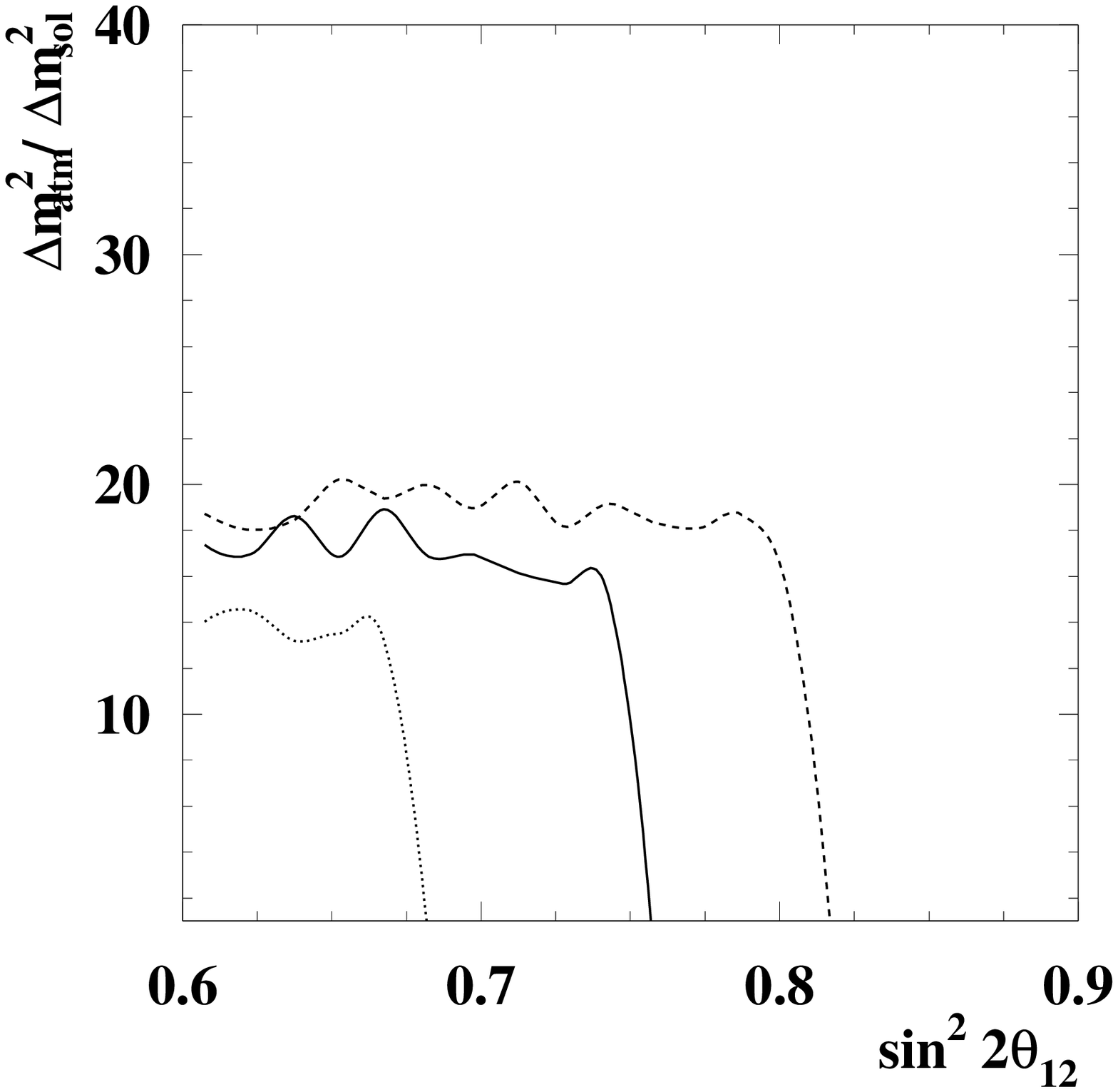}
   }
\caption{Maximum of the ratio $R_{a/s}$ as a function of the
atmospheric mixing angle $\theta_{23}$ (left)
and solar angle (right), for $\delta_{CKM} = 80^o$ (dotted line),
$90^o$ (solid) and $100^o$ (dashed). Additional cuts are
$\sin^2 2\theta_{12} > 0.7$ (left) and $\sin^2 2\theta_{23} > 0.9$ (right).}
\label{fig_23_del}
\end{figure}

However, the results discussed above were obtained for a value of the
CKM phase $\delta = 90 \deg$ which is too large compared with the measured
value ($\delta_{CKM} = 59 \pm 13 \deg$ from PDG \cite{pdg}).
As argued in section III (and indeed noted by previous analysis)
there is a strong dependence  on the goodness of the fit on the value
of $\delta_{CKM}$, with larger values giving better fits.
We show this dependence in Fig. \ref{fig_23_del}. The  parameters
are the same as in Fig \ref{fig_90_18}
($m_b(M_Z) = 3.11$ GeV, $m_t = 181$GeV, $M_{SUSY} = 500$ GeV and
$\tan \beta = 50$), but the three lines correspond to different values
for $\delta_{CKM}$: $80^o$ (dotted line),  $90^o$ (solid) and
$100^o$ (dashed line). One can notice a rapid deterioration in the
goodness of the fit with decreased $\delta_{CKM}$.
 Thus,
for $\delta_{CKM} = 80 \deg$, the best fit to neutrino sector we find
(after scanning over the SUSY parameter space) is
$\sin^2 2\theta_{23} \simeq 0.88$, $\sin^2 2\theta_{12} \simeq 0.7$ and
$R_{a/s} \simeq 18$.

For purposes of illustration, we give a fit obtained for a type--II dominant
case for $\delta_{CKM} = 80 \deg$, $m_b(M_Z) = 3.11$ GeV,
$M_t =$ 181 GeV, $\tan \beta = 55$ and $M_{SUSY} = 1$ TeV.
The $s, c$ quark masses at low scale
are $m_s(M_Z) = 0.074$ GeV, $m_c(M_Z) = 0.83$ GeV.
 Then the values of the quark and lepton masses at
GUT scale are (in GeV):

\bea{num_tii}
\begin{array}{rlrlrl}
m_u &= 0.0008185 \; &  m_c & =0.3772  & m_t & =139.876  \\
m_d & =0.0015588 \; &  m_s & =0.03554   &  m_b & =2.3547   \\
m_e & =0.000525 \; & m_\mu & =0.1107 \ \  & m_\tau & =2.420 \\
s_{12} & = 0.225 &  s_{23} & = 0.0297 & s_{13} & = 0.00384 \\
& &  \delta_{CKM} & = 1.4 \ . & &
\end{array}
\eea

 The values
of GUT scale phases (in radians) and $a,b$ parameters are given by:
\bea{num2_tii}
\begin{array}{rlrlrl}
 a_u & = -0.4689  &    a_c & = -1.0869 \ &  a_t & = 3.0928  \\
b_d & = 2.6063 &     b_s & = 2.2916 &  b_b & = 0.  \\
  a & = 0.09093 \ &  b & = 4.423 &  \sigma & = 3.577 \ .
 \end{array}
\eea
The running factors for the neutrino mass matrix are
$r_{22} = 1.09, r_{23} = 1.18$.

One then obtains for the neutrino parameters at low scale:
$$\Delta m^2_{23}/\Delta m^2_{12} \simeq 18 \ , \
\sin^2 2 \theta_{12}  \simeq 0.7 \ , \ \sin ^2 2\theta_{23} \simeq 0.88\
, \ \sin ^2 2\theta_{13} \simeq 0.094 \ .$$
The elements of the diagonal
neutrino mass matrix (masses and Majorana phases) are
$$ m_{\nu i} \ \simeq \ \{0.0016 \exp(0.27i)\ ,
\ 0.011\exp(-2.86i)\ ,\ 0.048  \}
$$
in eV.

The Dirac phase appearing in the MNS matrix is $\phi_D = -0.007$rad,
and one evaluates the effective neutrino mass for the neutrinoless double
beta decay process to be
$$ | \sum U_{e i}^2 m_{\nu i}| \ \simeq \ 0.01 \ \hbox{eV} \ .
$$

\subsection{Type--I seesaw}

The fact that in type--II seesaw one can obtain large mixing in the 23
sector is due to a lucky coincidence: the type--II neutrino mass matrix
being written as a sum of two hierarchical matrices ($M_l$ and $M_d$),
the most natural form for the neutrino mass matrix is also hierarchical.
However, since 33 elements of both matrices $M_l$ and $M_d$
are roughly of the same magnitude, by choosing the relative phase between
the two to be close to $\pi$, one can get a neutrino mass matrix of the
form suited to explain large mixing in the 2-3 sector.

The question arises then if such a coincidence happens for the type--I
seesaw neutrino mass matrix. To see that, let's write the Dirac neutrino
mass matrix in the following form:
$$ M_\nu^D \ = \ {b e^{i \sig } + 2 \over a}
\left[ \tilde{M_R} + {b - e^{-i\sig} \over b e^{i\sig } +2} M_d \right]
\sim \tilde{M_R} + \tilde{M_d}
$$
where $\tilde{M_R}$ is the scaled right-handed neutrino mass
matrix $\tilde{M_R} = M_l - M_d e^{i\sig}$ and $\tilde{M_d}$ is a
rescaled down-type quark diagonal mass matrix (the scaling factor
in this later case  $\zeta = (b - e^{-i\sig}) / (b e^{i\sig } +2)$
is close to unity, since $b$ is roughly of order 10). Then the
type--I seesaw neutrino mass matrix would be: \beq{typeI_dec}
 M_{\nu I} \ = \ M_\nu^D M_R^{-1} M_\nu^D \ \sim \ \tilde{M_R}
+ 2 \tilde{M_d} + \tilde{M_d} \tilde{M_R}^{-1} \tilde{M_d} \ .
\eeq

Now, for most values of the phase $\sig$, $\tilde{M_R}$
is hierarchical,
therefore so is $\tilde{M_R}^{-1}$ , therefore
the type--I neutrino mass matrix is the sum of three hierarchical matrices
($\tilde{M_d}$ being diagonal). So it is not surprising that
for most values of the phase $\sig$ $ M_{\nu I}$ is also hierarchical. What is
remarkable is that there are some values of  $\sig$ for which
the type--I seesaw mass matrix has a large mixing in the 2-3 sector, and
moreover, this happens for the same values of $\sig$ as in the case
when the type--II mass matrix  is non-hierarchical (that is, $\sig$ close
 to $\pi$).

In order to see this let us consider the magnitude and the phase
of the 33 elements (the largest ones) in the three terms on the
right-hand side of Eq. (\ref{typeI_dec}). If $\sigma$ is not close
to $\pi$, the magnitude of $\tilde{M_R}_{33}$ is of order $m_b$,
with varying phase ($\phi$ in Fig. \ref{fig_1} in the first
quadrant); the magnitude of  $(\tilde{M_d})_{33}$ is also of order
$m_b$, and the phase $\sim -\sig$. For the last term, we make use
of the fact that $\tilde{M_R}$ being hierarchical,
$(\tilde{M_R}^{-1})_{33} \simeq 1/(\tilde{M_R})_{33} \simeq
1/m_b$; then $\tilde{M_d} \tilde{M_R}^{-1} \tilde{M_d} \simeq
m_b$, with a phase close to $-2 \sig$. We see then that for most
values of $\sig$ $(M_{\nu I})_{33}$ is of order $m_b$, while the
off-diagonal elements are small. However, for $\sig \sim \pi$, the
cancellation in the 33 element of $\tilde{M_R}$ is matched by a
cancellation between the 33 elements of the $2 \tilde{M_d}$ and
$\tilde{M_d} \tilde{M_R}^{-1} \tilde{M_d}$ terms from Eq.
(\ref{typeI_dec}) (since the relative phase between these is also
$\sig$), thus leading to a non-hierarchical form for the type--I
seesaw neutrino mass matrix.

The fine-tuning between different contributions to the neutrino
mass matrix is thus a little bit more involved in the type--I
seesaw case compared to the type--II seesaw, but it can still lead
to large mixing in the 2-3 sector. Moreover, since the
correlations between the input parameters and the neutrino mass
matrix elements are not so strong, most of the constraints
discussed in the above section do not hold (for example, $m_b$
does not have to be necessarily very close to $m_{\tau}$). This
may lead one to believe that it is possible to obtain a better fit
for the neutrino sector in type--I models, and we found that in
fact this is the case.

\begin{figure}[t!] 
\centerline{
   \includegraphics[height=3.in]{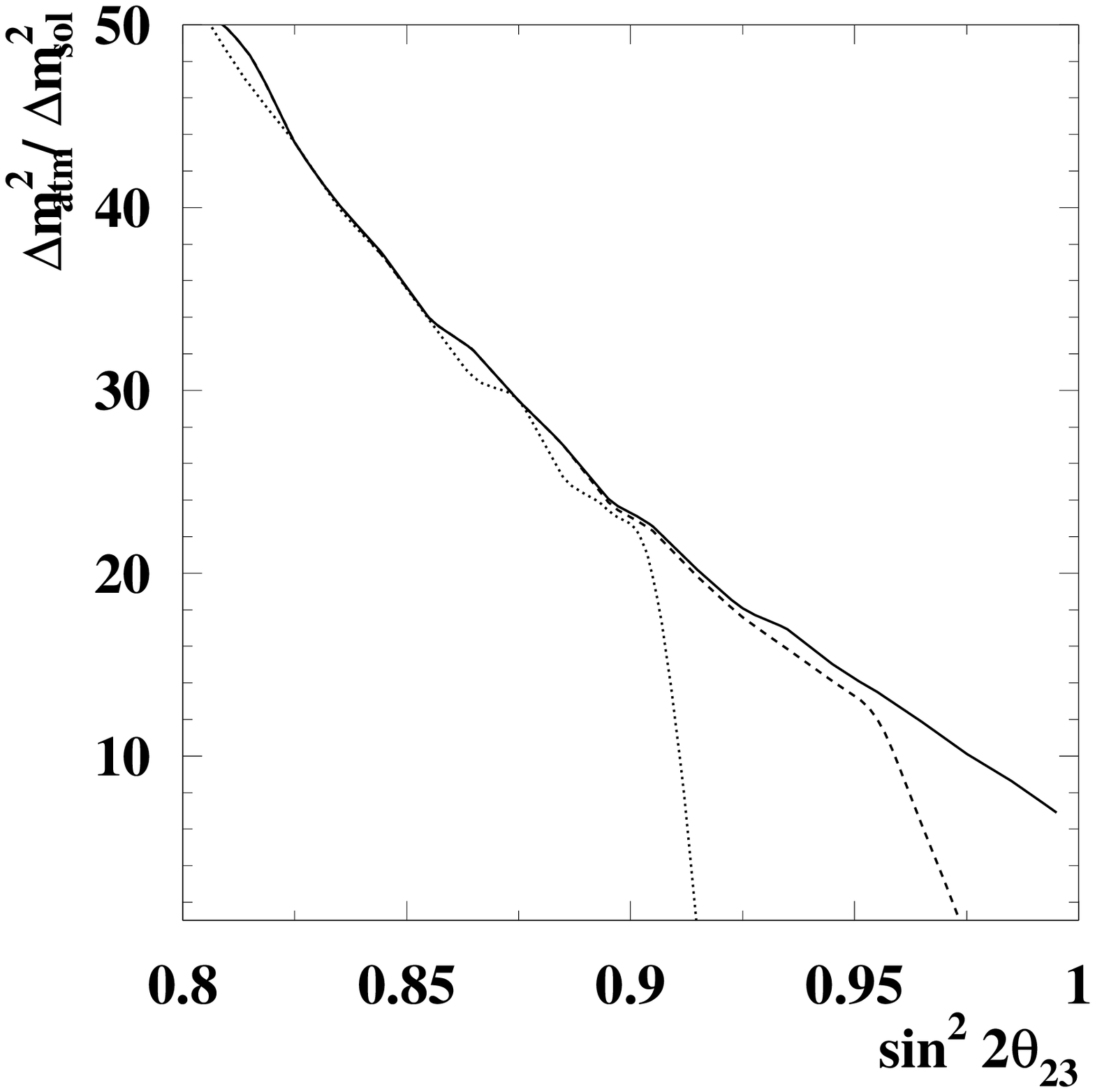}
   \includegraphics[height=3.in]{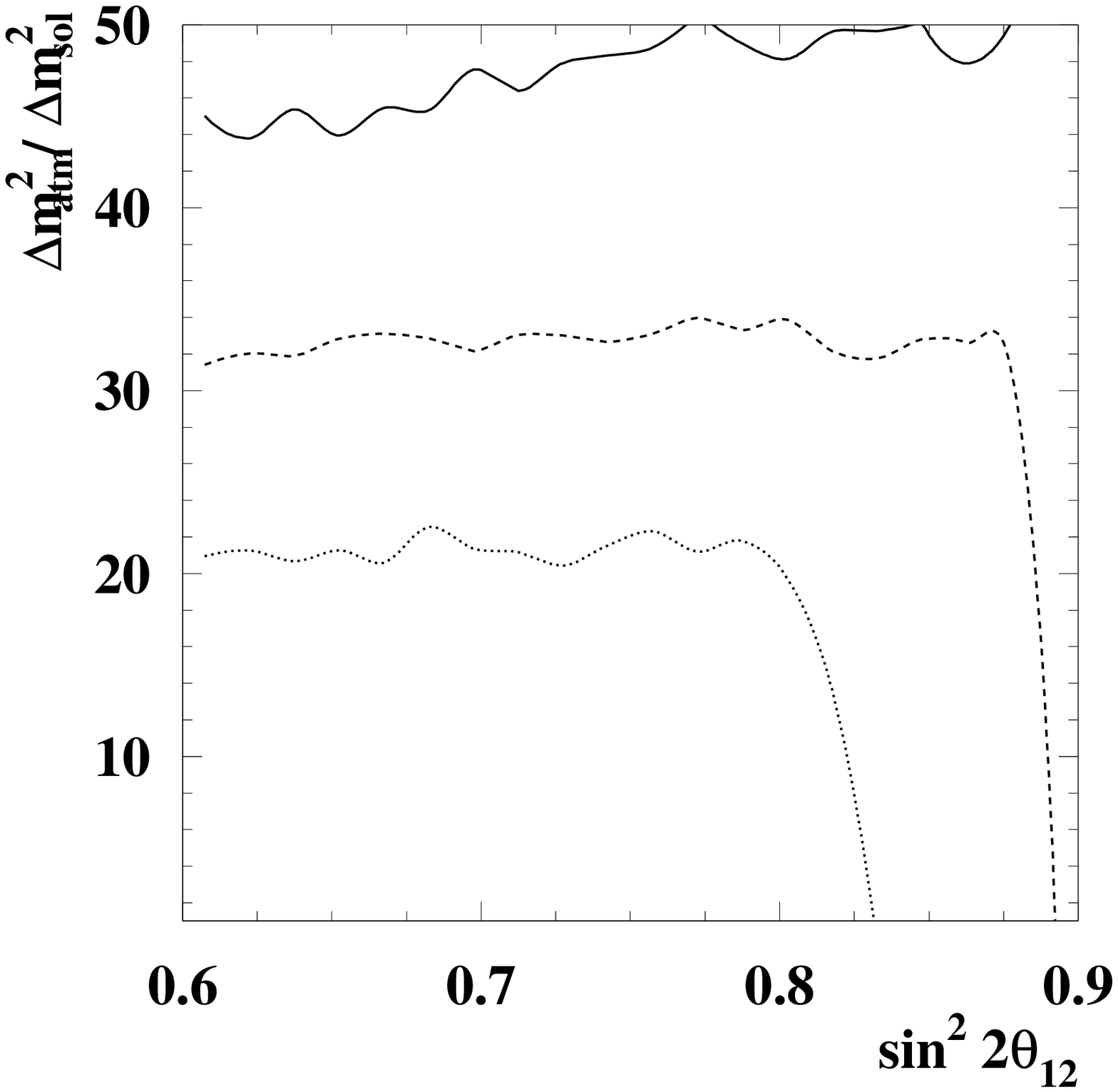}
   }
\caption{Left: maximum of the atmospheric/solar mass splitting ratio
$R_{a/s}$ as a function of the
atmospheric mixing angle $\theta_{23}$, with cuts on the solar angle
$\sin^2 2\theta_{12} > 0.8$ (dotted line),  $ 0.7$ (dashed)
and no cut (solid).
 Right: maximum of $R_{a/s}$ as a function of the
solar mixing angle $\theta_{12}$, with cuts on the atmospheric angle
$\sin^2 2\theta_{23} > 0.9$ (dotted line),  $ 0.85$ (dashed)
and $ 0.8$ (solid).}
\label{fig_ii}
\end{figure}

For example, we show in Fig \ref{fig_ii}(left) the
maximum
atmospheric/solar mass splitting ratio $R_{a/s}$
as a function of the atmosperic mixing angle $\theta_{atm} = \theta_{23}$ -
 with cuts on the
solar mixing angle: $\sin^2 2\theta_{12} > 0.8$ (dotted), $\sin^2
2\theta_{12} > 0.7$ (dashed) and no cut (solid). In the left panel
we show maximum $R_{a/s}$ as a function of the solar angle for
$\sin^2 2\theta_{23} > 0.9$ (dotted),  $\sin^2 2\theta_{23} >
0.85$ (dashed) and $\sin^2 2\theta_{23} > 0.8$ (solid line). This
figure is obtained for values $m_b(M_Z) = 3.0$ GeV, $m_t =
174$GeV, $M_{SUSY} = 500$GeV and $\tan \beta = 40$, while the CKM
phase is allowed to vary between 60 and 70 deg. We see that it is
possible to obtain a large atmosperic/solar mass splitting ratio
for values of the atmospheric and solar mixings consisted with
experimental constraints.

\begin{figure}[t!] 
\centerline{
   \includegraphics[height=3.in]{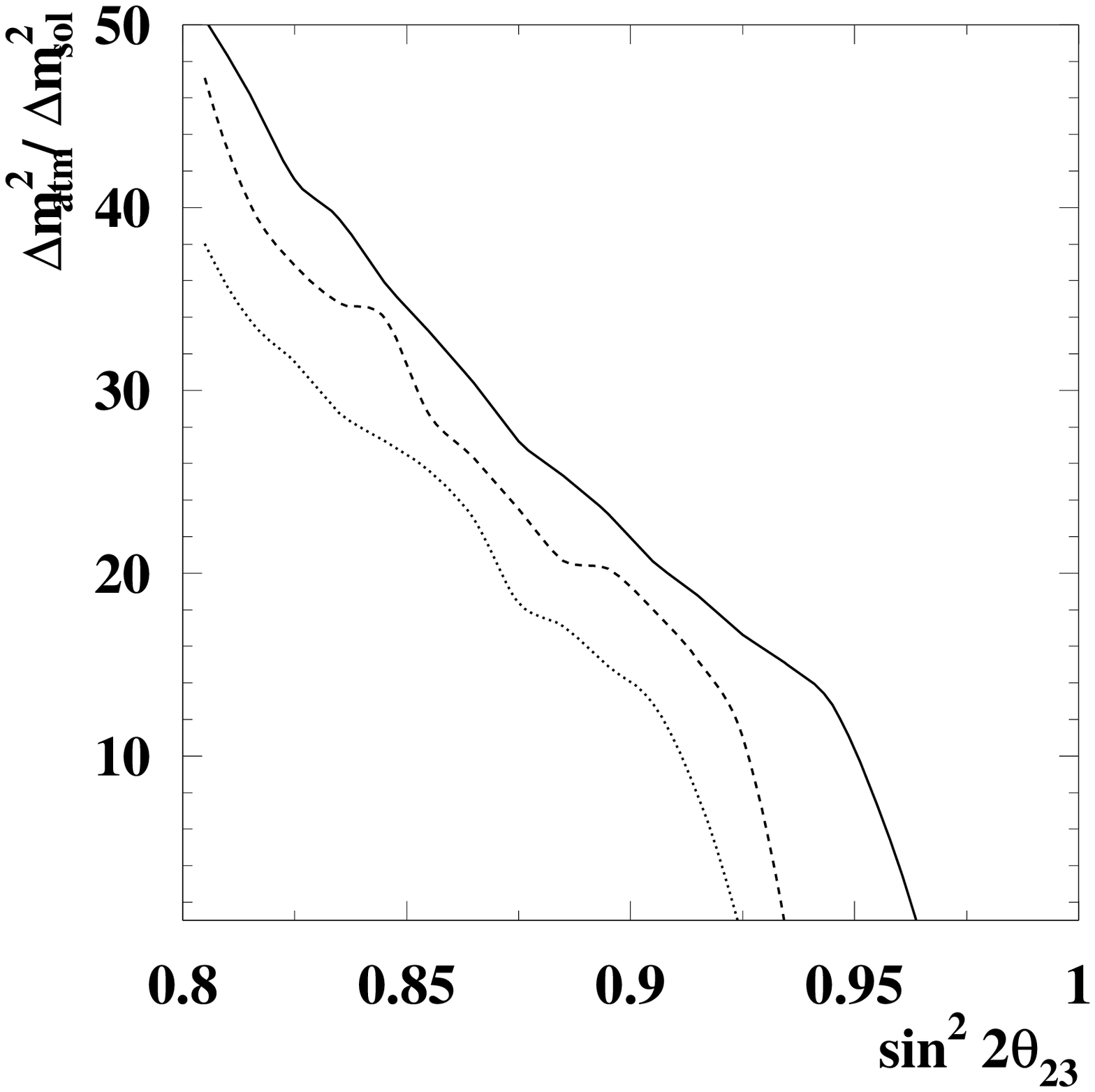}
   \includegraphics[height=3.in]{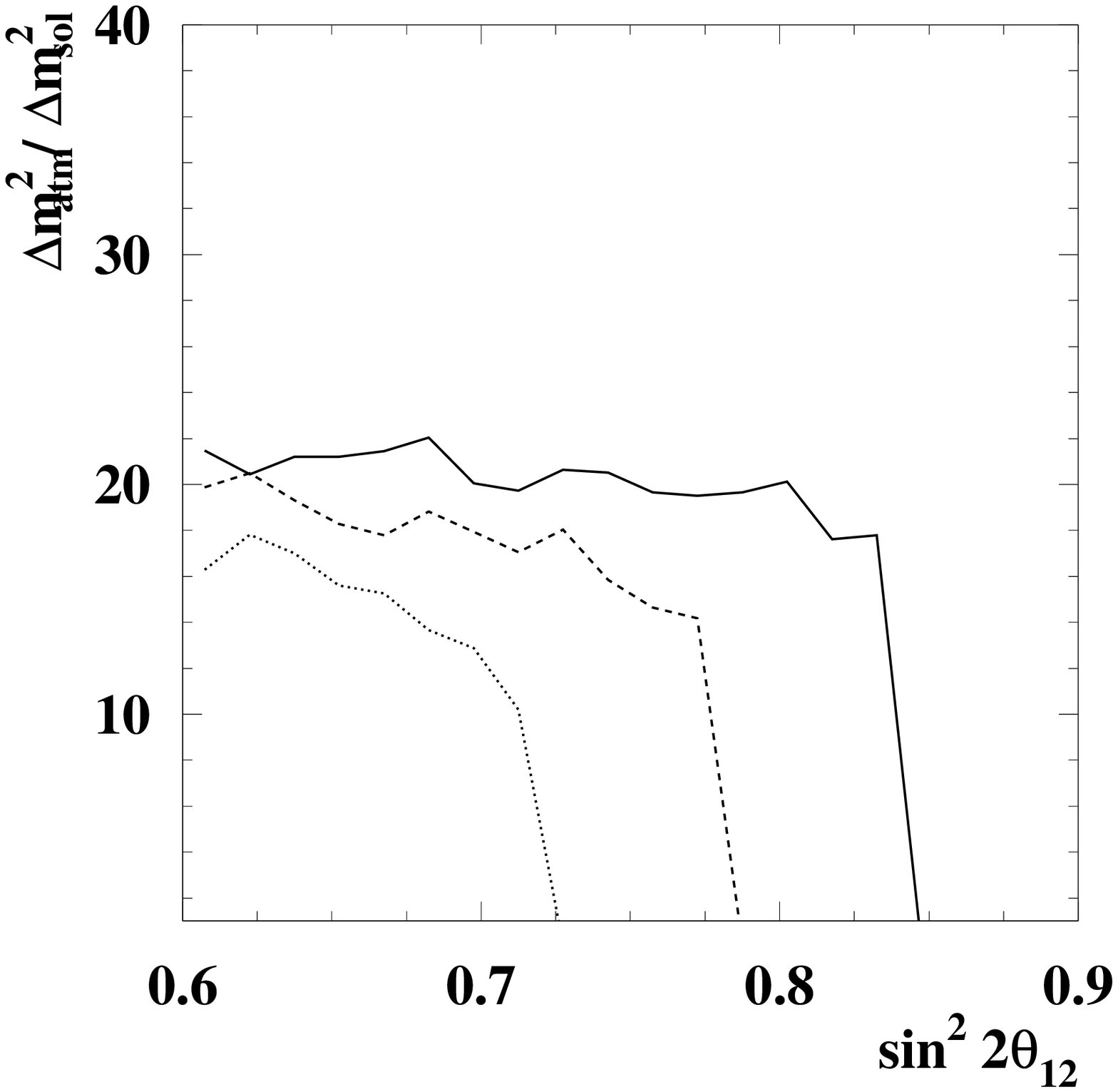}
   }
\caption{Maximum of the ratio $R_{a/s}$ as a function of the
atmospheric mixing angle $\theta_{23}$ (left) and solar angle $\theta_{12}$
(right), for $\tan \beta = 20$ (solid line),
45 (dashed) and 55 (dotted). Additional cuts are
$\sin^2 2\theta_{12} > 0.7$ (left) and $\sin^2 2\theta_{23} > 0.9$ (right).}
\label{fig_iia}
\end{figure}

How do these results change if we modify the SUSY parameters $\tan
\beta$ and $M_{SUSY}$, and/or the $M_Z$ scale masses? We find that
one can get good fits for values of the parameter $\eta_{b/\tau} =
m_b/m_{\tau}(M_{GUT})$ between 0.83 and 0.9. For values of
$m_b(M_Z) = 3.0$ GeV, $m_t = 174$ GeV, $M_{SUSY} = 500$ GeV, this
means that $\tan \beta$ can vary between 10 and 55. However, if
one increases $m_b$ or $m_t$, generally $\eta_{b/\tau}$ will
increase (there is also a slight decrease with increasing
$M_{SUSY}$, but less pronounced). Thus, for $m_b(M_Z) = 3.11$ GeV,
$m_t = 174$ GeV, one has to take $\tan \beta$ roughly between 10
and 45 in order to get a good neutrino result. Fig. \ref{fig_iia}
illustrates this behavior: the three lines correspond to values $
\tan \beta = 20$ (solid), 45 (dashed) and 55 (dotted), for
$m_b(M_Z) = 3.11$ GeV, $m_t = 174$ GeV and $M_{SUSY} = 500$ GeV
the corresponding values for $\eta_{b/\tau}$ are 0.86, 0.9 and
0.96. As noted above, the fit worsens dramatically for $\tan \beta
> 45$. $\delta_{CKM}$ is taken between $60$ and $70$ degrees here
also.

\begin{figure}[t!] 
\centerline{
   \includegraphics[height=3.in]{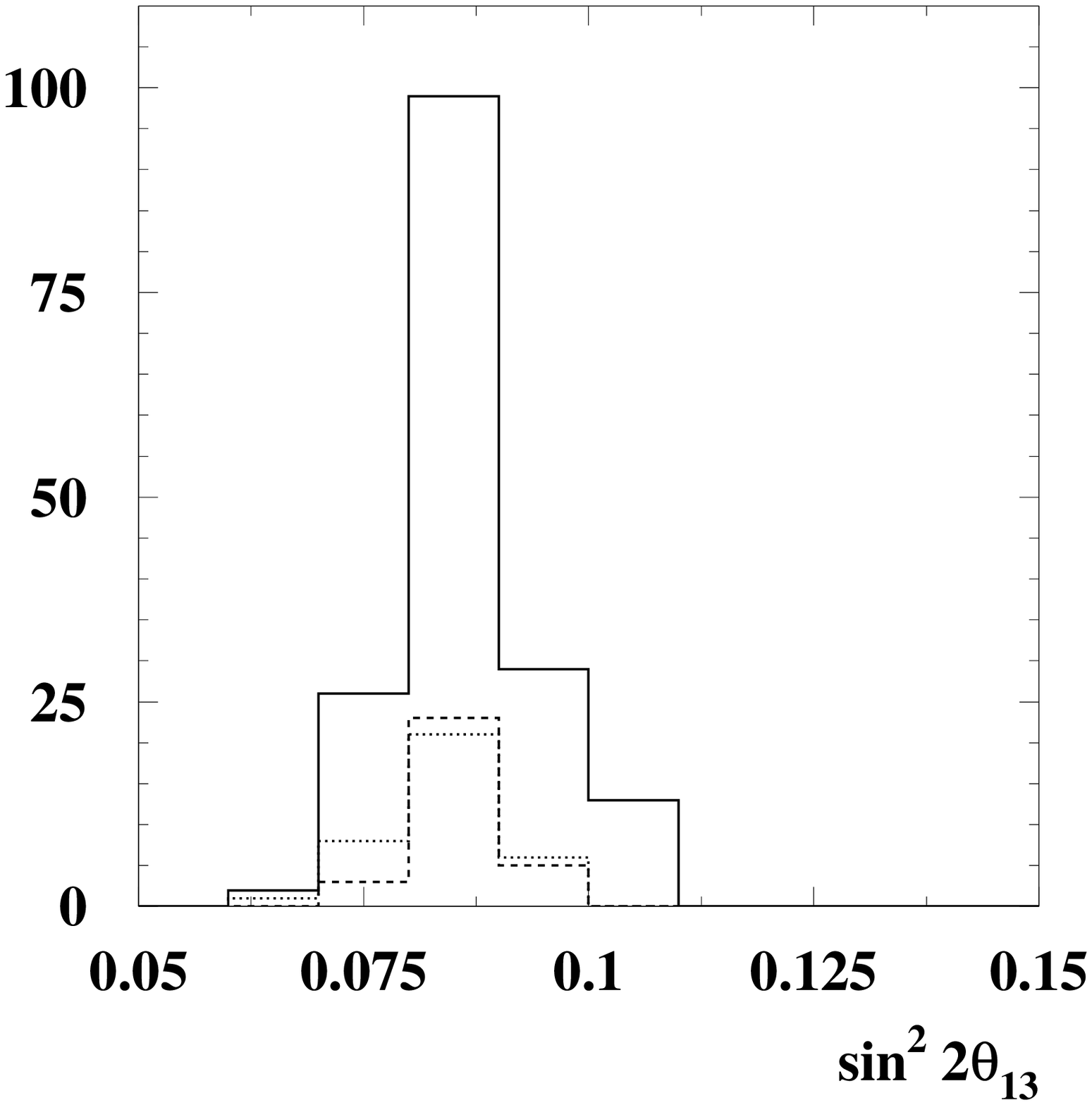}
   \includegraphics[height=3.in]{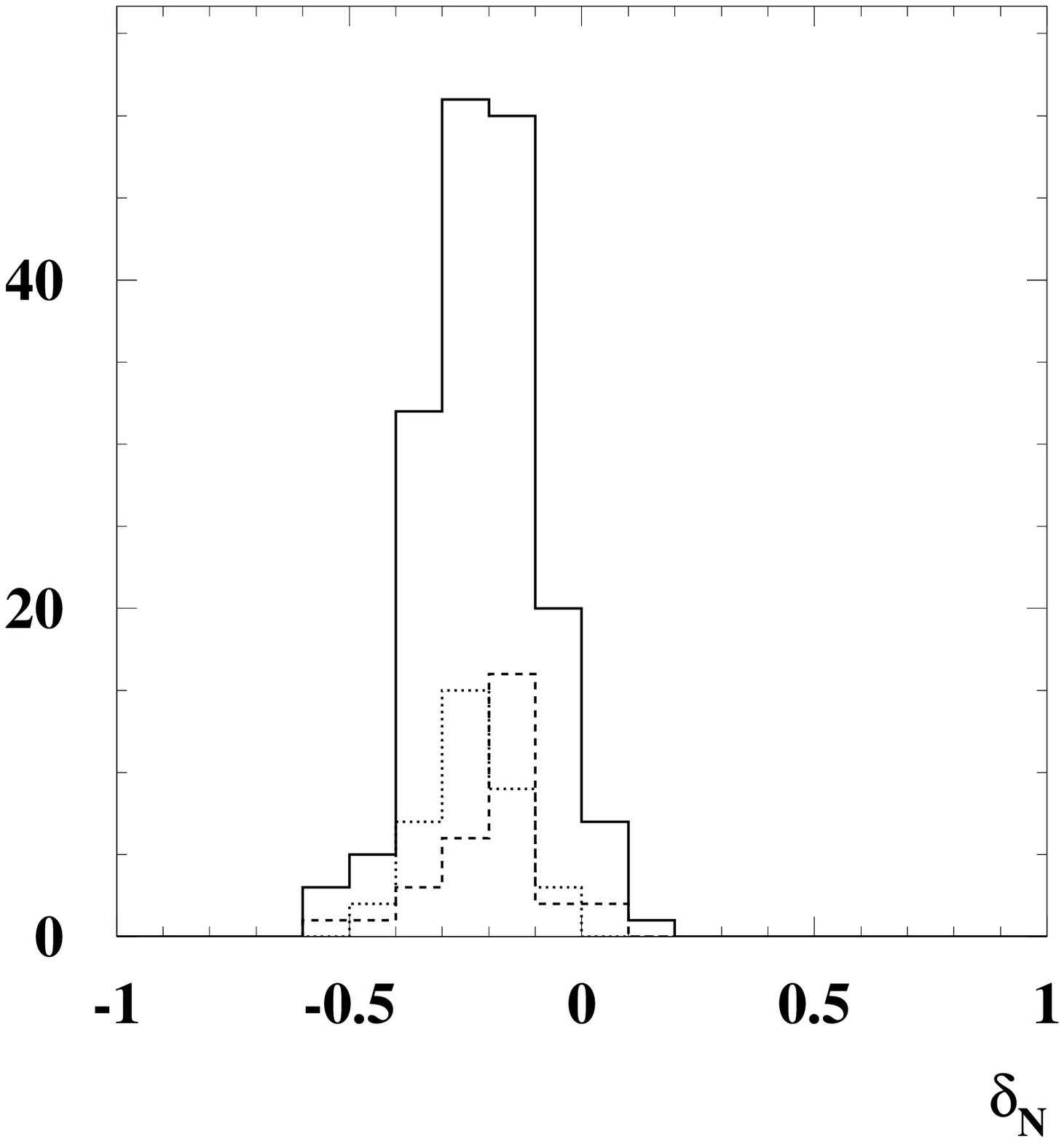}
   }
\caption{Distribution of values for the $\theta_{13}$ mixing
angle (left) and the Dirac phase in the neutrino mixing matrix $\delta_N$
(right)
 obtained after a scan of the parameter space. The three lines
correspond to different cuts on the other neutrino sector paramenters
(see text for details).  }
\label{fig_sin13}
\end{figure}

Finally, the $\theta_{13}$ mixing angle is found to lie in a range
$ 0.06 \lesssim \sin^2 2\theta_{13} \lesssim 0.11$, with
a preferred value $\sim$ 0.085 . We show in Fig. \ref{fig_sin13}
the distribution of values for the $\theta_{13}$ mixing angle
and the Dirac phase in the neutrino mixing matrix $\delta_N$ (in radians)
obtained for
a scan of parameter space with 2.95  GeV $\lesssim m_b(M_Z) \lesssim $ 3.05 GeV,
172  GeV $\lesssim m_t \lesssim $ 176 GeV, 500 GeV $\lesssim M_{SUSY}
 \lesssim $ 750 GeV, and 0.83 $\lesssim \eta_{b/\tau} \lesssim $ 0.9 .
The cuts on the other neutrino sector parameters are
$\sin^2 2\theta_{23} > 0.88$ (for all lines),
$\sin^2 2\theta_{12} > 0.7$ and $R_{a/s} > 18$ (solid line),
$\sin^2 2\theta_{12} > 0.75$ and $R_{a/s} > 18$ (dashed line),
$\sin^2 2\theta_{12} > 0.7$ and $R_{a/s} > 20$ (dotted line).
Note that the preferred value for $\theta_{13}$ is close to the experimental
limit.

\subsection{Mixed type}

We end by considering the scenario when both types of seesaw are present,
and give a non-negligible contribution. One can envision two
different situations: first when one contribution dominates, and
the other can be treated as a small perturbation, and second when the
two contributions are of roughly the same order of magnitude.

Let us first discuss the first case. It is obvious that the fits one obtains
when considering the `pure' case (either type--I or type--II) can only improve.
Indeed, by adding the other type contribution, we introduce another two
free parameters: the relative magnitude and phase of the subdominant
contribution relative to the dominant one. One can then obtain the
results for the `pure' case by allowing the relative magnitude go to zero.

Thus there appears an interesting question. Is it possible to
improve the fit for the type--II scenario to a level which is
compatible with experimental data? Unfortunately, the answer seems
to be no. Our numerical simulations show that by adding type--I
contribution as a small perturbation, one obtains a small
improvement over the `pure' type--II case, but not a significant
one.

This might seems counterintuitive at first glance. It would be
reasonable to assume that if one takes the best result for a
type--II fit and then adds another small quantity parametrized in
terms of a free magnitude and phase, one should be able to obtain
a somewhat better fit. The reason why this does not happen is that
the type--I seesaw perturbation will modify in first order only
the 33 neutrino matrix element\footnote{ It is easy to see that in
the sum of last two terms on the right-hand side of Eq.
(\ref{typeI_dec}) the 33 element is dominant if the type--II 23
mixing is close to maximal.}. However, as we saw from our analysis
of type--II case, the goodness of the fit is determined mainly by
the 22 and 23 matrix elements. That is, in type--II case, one
generally has the freedom to easily modify the 33 element (by
changing the phase $\sigma$ or the ratio $m_b/m_{\tau}$) so in a
sense for a best fit $m_{33}$ already has its optimal value, and
adding a small perturbation to it does not buy any additional
freedom. At most what one can achieve is to take some fit which is
not optimal and make it better by adding the type--I perturbation.

Let us now consider the case when the contributions from  the two
types of seesaw are of comparable magnitude. In this case, the
neutrino mass matrix can be thought of as having the form
displayed in Eq. (\ref{typeI_dec}), with the first term
$\tilde{M_R}$ (proportional to the type--II seesaw contribution)
enhanced or diminished, depending on the relative phase between
the two contributions. We find that if the type--II term is close
to zero (that is, both contributions have about the same magnitude
and the relative phase is close to $\pi$) then we can again get
good fits for the neutrino sector. In particular, in this case we
can obtain truly maximal mixing in the atmospheric sector ($\sin^2
2 \theta_{23} \simeq 1$) with large enough mass splitting, and
large enough solar angle. Also, obtaining these results does not
require that we use a specific range of (low scale) parameters -
like large $\tan \beta$, or large $m_b$. Generally, as long as one
can fit for the lepton-quark sector, one can obtain a fit for the
neutrino sector, too.

Here is a particular fit obtained in a mixed case. We took
$\delta_{CKM} = 60 \deg$, $m_b(M_Z) = 3.0$ GeV, $M_t =$ 181 GeV,
$\tan \beta = 45$ and $M_{SUSY} = 500$ GeV. The $s, c$ quark
masses at low scale are $m_s(M_Z) = 0.071$ GeV, $m_c(M_Z) = 0.79$
GeV.
 Then the values of the quark and lepton masses at
GUT scale are (in GeV):

\bea{num_comb}
\begin{array}{rlrlrl}
m_u &= 0.0009440 \; &  m_c & =0.3958  & m_t & =143.23  \\
m_d & =0.001451 \; &  m_s & =0.02991   &  m_b & =1.6903   \\
m_e & =0.00047 \; & m_\mu & =0.0992 \ \  & m_\tau & = 1.910 \\
s_{12} & = 0.220 &  s_{23} & = 0.0320 & s_{13} & = 0.00249 \\
& &  \delta_{CKM} & = 1.05 \ . & &
\end{array}
\eea

 The values
of GUT scale phases (in radians) and $a,b$ parameters are given by:
\bea{num2_comb}
\begin{array}{rlrlrl}
 a_u & = -1.2527  &    a_c & = -0.39256 \ &  a_t & = 3.07385  \\
b_d & = 3.0896 &     b_s & = -3.1367 &  b_b & = 0.  \\
  a & = 0.089727 \ &  b & = 6.6365 &  \sigma & = 3.1505 \ .
 \end{array}
\eea
The running factors for the neutrino mass matrix are
$r_{22} = 1.04, r_{23} = 1.09$. We add the two contributions:
$$ M_\nu \ \sim \ M_{\nu II} {(M_{\nu I})_{33} \over (M_{\nu II})_{33}}
\ +  \ M_{\nu I} \ r e^{i \phi}   \  ,$$
with relative magnitude $r = 0.9813$ and phase $ \phi = 3.200 $rad.

One then obtains for the neutrino parameters at low scale:
$$\Delta m^2_{23}/\Delta m^2_{12} \simeq 28 \ , \
\sin^2 2 \theta_{12}  \simeq 0.85 \ , \ \sin ^2 2\theta_{23}
\simeq 0.98\ , \ \sin ^2 2\theta_{13} \simeq 0.05 \ .$$ The
elements of the diagonal neutrino mass matrix (masses in eV and
Majorana phases) are
$$ m_{\nu i} \ \simeq \ \{0.0028 \exp(0.026i)\ ,
\ 0.0093\exp(-3.07i)\ ,\ 0.0478  \}~.
$$
 The Dirac phase appearing in the MNS matrix is $\phi_D =
-0.3$rad, and one evaluates the effective neutrino mass for the
neutrinoless double beta decay process to be
$$ | \sum U_{e i}^2 m_{\nu i}| \ \simeq \ 0.009 \ \hbox{eV} \ .
$$

\section{Conclusions}

An $SO(10)$ model with only one {\bf 10} and one $\overline{\bf
126}$ Higgs multiplets coupling to fermions provides an appealing
candidate for a unified theory at large scales. The number of free
parameters in such a model is smaller than the number of free
parameters in the Standard Model, thus giving the theory some
predictive power. In particular, there appears the possibility
that large mixing in the neutrino sector can be understood as a
consequence of $b - \tau$ unification at GUT scale.

We revisit here the analysis of the minimal $SO(10)$ model and its
implications for the neutrino sector. Our work differs from
previous works in this area in several aspects. First, we consider
the most general formulation of the model, with all the CP phases
taken into consideration. Second, we use a new method for fitting
the GUT scale parameters (Yukawa couplings) to the low scale
masses and mixing angles in the quark and lepton sector. The
running of the Yukawa couplings from low scale to unification
scale is also taken into account, as well as dependence upon SUSY
parameters like tan$\beta$ and $M_{SUSY}$. We also analyze all
possible cases for the neutrino mass generation, namely, when
either the type--I or type--II seesaw mechanism dominates, or the
case when both contributions are roughly of the same magnitude.

Our results are as follows. For the type--II seesaw case, we find
that the requirement for close to maximal mixing in the 2-3
sector, together with the large hierarchy between atmospheric and
solar mass splittings, pushes the CKM phase to large values
(reasonably good fits can be found for $\delta_{CKM} > 100^o$).
This is in agreement with previous results. Interestingly, we also
find the requirement that the solar mixing angle is large imposes
significant constraints. Better fits can be found at large
tan$\beta$, and large values of $m_b(M_Z)$; however, the results
obtained for values of $\delta_{CKM}$ in the first quadrant are at
most marginally in agreement with experimental data.

More interesting is the case when the type--I seesaw mechanism
dominates. Here, contrary to previous analyses, we find that it is
possible to obtain good fits for the neutrino sector for values of
$\delta_{CKM}$ as low as $50^o$, which includes the range
consistent  with experimental results. Actually, unlike the
type--II case, it seems that the goodness of the fit is not very
dependent on  $\delta_{CKM}$; the relevant parameters seem to be
the values of $m_c$ and $m_s$ at $M_Z$ scale, with preference for
larger values for $m_c$ and lower values for $m_s$. The solar
angle also imposes significant constraints on the parameter space,
but in this case it is possible to obtain good fits for a larger
range of tan$\beta$, $m_b(M_Z)$ and $M_t$.

It is also interesting to note that, for values of the phase
$\sig$ appearing in the neutrino sector close to $\pi$, the
type--I and type--II neutrino mass matrices are roughly
proportional. This means that if the type--I mass matrix is
non-hierarchical (thus leading to large mixing in the 2-3 sector),
so will the type--II mass matrix be. Hence, if one considers the
case when both contributions from type--I and type--II must be
taken into account, one sees that the dominant contribution will
determine the type of the fit. However, a special case is when the
two contributions are roughly of the same magnitude, and their
relative phase is close to $\pi$. In this case they cancel each
other, and the resulting neutrino mass matrix is generally
non-hierarchical. One can obtain very good fits for the neutrino
sector in this case, too.

\section*{Acknowledgments}

This work is supported in part by U.S. Department of Energy under grants
\#DE-FG02-85ER40231, \#DE-FG02-04ER46140 and \#DE-FG02-04ER41306.

\end{document}